\documentclass[a4paper,prd,preprintnumbers,twocolumn,superscriptaddress,nofootinbib,amsmath,amssymb]{revtex4-2}
\usepackage{hyperref}
\hypersetup{
  colorlinks=true,        
  linkcolor=blue,         
  citecolor=cyan,         
}

\usepackage{graphicx,float}
\usepackage{mathrsfs,array,multirow}
\usepackage{amstext}
\usepackage{subfigure}
\usepackage{dcolumn}
\usepackage{bm}
\usepackage{color}
\usepackage{enumitem}
\usepackage{amsmath}
\usepackage{amssymb}
\usepackage{subcaption}

\renewcommand{\arraystretch}{1.5}

\begin{document}

\title{Probing scalarized wormholes through quasi-periodic oscillations and spinning particle dynamics}

\author{Asalkhon Alimova}
\email{asalxon2197@gmail.com}
\affiliation{Institute of Fundamental and Applied Research, National Research University TIIAME, Kori Niyoziy 39, Tashkent 100000, Uzbekistan} 
\affiliation{University of Tashkent for Applied Sciences, Str. Gavhar 1, Tashkent 100149, Uzbekistan}

\author{Akbar Davlataliev}
\email{akbar@astrin.uz}
\affiliation{School of Physics, Harbin Institute of Technology, Harbin 150001, People’s Republic of China}
\affiliation{New Uzbekistan University, Movarounnahr str. 1, Tashkent 100000, Uzbekistan}

\author{Farruh~Atamurotov}
\email{atamurotov@yahoo.com}
\affiliation{Kimyo International University in Tashkent, Shota Rustaveli str. 156, Tashkent 100121, Uzbekistan}
\affiliation{Research Center of Astrophysics and Cosmology, Khazar University, 41 Mehseti Street, Baku AZ1096, Azerbaijan}

\author{Ahmadjon~Abdujabbarov}
\email{ahmadjon@astrin.uz}
\affiliation{School of Physics, Harbin Institute of Technology, Harbin 150001, People’s Republic of China}
\affiliation{National University of Uzbekistan, Tashkent 100174, Uzbekistan}

\author{Phongpichit Channuie}
\email{phongpichit.ch@mail.wu.ac.th (\textcolor{red}{Corresponding Author})}
\affiliation{School of Science, Walailak University, Nakhon Si Thammarat, 80160, Thailand}
\affiliation{College of Graduate Studies, Walailak University, Nakhon Si Thammarat, 80160, Thailand}

\author{Chengxun Yuan}
\email{yuancx@hit.edu.cn}
\affiliation{School of Physics, Harbin Institute of Technology, Harbin 150001, People’s Republic of China}

\begin{abstract}

We investigate the dynamics of test particles in a three-parameter scalarized wormhole spacetime within Einstein-scalar field theory. For spinless particles, we derive the orbital and epicyclic frequencies and compute twin-peak QPO spectra using the ER3 and ER4 resonance models. The scalar coupling parameter $g_s$  shifts the innermost stable circular orbit to larger radii and systematically modifies the characteristic 3:2 resonance condition. Extending to spinning particles via the Mathisson-Papapetrou-Dixon formalism, we find that spin-curvature coupling significantly alters the effective potential and innermost stable circular orbit parameters. The maximum physically admissible spin increases monotonically with the scalar coupling. Analysis of particle collisions near the wormhole throat reveals that both scalar coupling and relative spin orientation determine collision energetics, with anti-aligned spin configurations producing substantially higher energies. Our results suggest that the combined effects of scalar coupling and spin-curvature interaction leave distinct imprints on QPO frequencies and collision processes, potentially providing observable signatures for distinguishing scalarized wormholes from standard black holes.

\textbf{Keywords:} Wormholes, scalar fields, quasi-periodic oscillations, Spining Particles.
\end{abstract}

\maketitle
\section{Introduction}

The general theory of relativity (GR) has provided a remarkably successful description of gravity on astrophysical and cosmological scales \cite{Will2014}. However, despite its successes, GR faces profound challenges, particularly in the ultraviolet regime, where quantum effects become important, and in the context of singularities that plague classical solutions like black holes \cite{Penrose1965,Hawking1970}. The presence of spacetime singularities, where curvature invariants diverge and physical laws break down, is widely considered an indication that GR is an incomplete theory. This has motivated extensive research into regular singularity-free alternatives to black holes \cite{Bardeen1968,AyonBeato2000,Hayward2006}. Among these, wormholes that are hypothetic topological bridges connecting different regions of spacetime stand out as one of the most intriguing solutions \cite{Morris1988,MorrisThorne1988,Visser1995}. A key attribute of many wormhole solutions is their ability to be geodesically complete, thereby avoiding the singularities inherent in their black hole counterparts, making them valuable theoretical laboratories for exploring strong-field gravity beyond the standard paradigm \cite{Lobo2008,Cataldo2010}.

In the context of modified gravity theories, scalar fields naturally emerge as additional degrees of freedom \cite{BransDicke1961,Faraoni2004,Clifton2012}. They play a central role in many frameworks beyond-GR, including scalar-tensor theories, Horndeski gravity, and string theory \cite{Horndeski1974,Deffayet2011,Kobayashi2019}. These scalar fields can couple non-minimally to matter, leading to observable deviations from GR, particularly in the strong-field regime \cite{DamourEspositoFarese1992,Perivolaropoulos2010}. For instance, the presence of a scalar field has been shown to support traversable wormhole geometries without the need for exotic matter in some cases \cite{Bronnikov1973,Ellis1973,Volkov2021}, or to generate new classes of black hole solutions with hair, such as the three-parameter scalarized wormhole spacetime considered in this work \cite{Turimov2022}. This model, parameterized by the mass ($M$), throat radius ($r_0$), and scalar charge ($\sigma$), provides a concrete platform to study the interplay between scalar fields and spacetime structure.

Observational astrophysics offers a promising avenue for discriminating between GR and alternative theories by probing the environments of compact objects \cite{Psaltis2008,Johannsen2013,Berti2015}. One of the most powerful tools in this regard is the study of quasi-periodic oscillations (QPOs), which are observed in the X-ray light curves of accreting black holes and neutron stars in X-ray binary systems \cite{RemillardMcClintock2006,IngramMotta2014}. These high-frequency modulations, often appearing as twin-peak features, are thought to originate from the motion of hot plasma blobs in the inner regions of accretion disks \cite{StellaVietri1998,Abramowicz2005}. Their frequencies are closely related to the characteristic orbital and epicyclic frequencies of test particles moving in the strong gravitational field near the compact object \cite{Kato1990,AlievGaltsov1981}. This intimate connection makes QPOs an exceptionally sensitive probe of the underlying spacetime geometry \cite{Silbergleit1998,Shapiro2004,Torok2005}. By analyzing the frequencies and frequency ratios of QPOs, it is possible to infer the properties of the central object, such as its mass, spin, and the nature of the background gravitational field \cite{Stella1999,Kluzniak2005}.

Epicyclic resonance models, such as the ER3 and ER4 models, provide a robust framework for interpreting these twin-peak QPOs \cite{AbramowiczKluzniak2001,KluzniakAbramowicz2001}. In these models, the observed frequencies are attributed to resonant interactions between the radial ($\nu_r$) and vertical ($\nu_\theta$) epicyclic oscillations of matter in the accretion disk, and sometimes their combination with the orbital (Keplerian) frequency ($\nu_\phi$) \cite{Abramowicz2003,Rezzolla2003}. A particularly robust observational feature is the presence of a stable 3:2 frequency ratio in the high-frequency QPOs of several microquasars, often referred to as the ``resonance condition" \cite{McClintockRemillard2006,Belloni2006,Stiele2011}. Observations from sources such as GRO J1655--40, XTE J1550--564, and GRS 1915+105 have provided strong evidence for these characteristic frequency ratios \cite{Remillard2002,McClintock2003,Motta2014}. The exact location at which this ratio is realized is highly sensitive to the background metric. Consequently, measuring the QPO frequencies from a known system can, in principle, be used to test the validity of different gravity models and to constrain or rule out exotic spacetimes \cite{Bambi2013,Barausse2015,Krawczynski2018}. Recent studies have successfully extended QPO analyses to a wide variety of modified gravity black holes, including Schwarzschild-like, noncommutative, metric-Palatini, Horndeski, and asymptotically safe models, further demonstrating the capability of orbital and epicyclic frequencies to probe the underlying spacetime geometry~\cite{Davlataliev:2024smq,Mustafa:2024cfe,Ghorani:2024ufk,Naseer:2025jtt,Mustafa:2025cou,Ashraf:2025elg}.

In this work, we investigate the dynamics of test particles in a three-parameter, scalarized wormhole spacetime. We aim to explore the effects of the scalar field, characterized by the coupling parameter $g_s$, on the dynamics of both spinless and spinning massive particles. The primary objectives are (i) to derive the fundamental frequencies of epicyclic motion for spinless particles, (ii) to calculate the resulting twin-peak QPO frequencies within the ER3 and ER4 resonance models, and (iii) to analyze the impact of the wormhole parameters on these observables. Furthermore, we extend our study to include spinning particles, where we investigate how spin--curvature coupling, mediated by the scalar field, modifies the properties of stable circular orbits and the efficiency of high-energy particle collisions near the wormhole throat \cite{MPDformalism,Semerak1999,Deriglazov2017}. In recent years, spinning particle dynamics has been extensively investigated in a variety of modified gravity black-hole spacetimes, including quantum-corrected, asymptotically safe, Reissner–Nordström-like, and hairy black holes, demonstrating that spin–curvature coupling provides a sensitive probe of the underlying spacetime geometry~\cite{Rakhimova:2024hzt,Umarov:2025wzm,Mannobova:2025uqf,Abdukayumova:2025ztr,Abdukayumova:2026zir}. Our analysis reveals that the scalar coupling and the wormhole geometry leave distinct imprints on the fundamental frequencies and QPO spectra, suggesting that these observables could serve as a valuable probe for distinguishing scalarized wormholes from standard black holes and other exotic compact objects \cite{Kunz2015,Sotiriou2015,Minamitsuji2016}.

This paper is organized as follows. In Sect.~\ref{sec2}, we present the Einstein--scalar field--massive particle system and the equations of motion. Sect.~\ref{sec3} describes the background wormhole spacetime and the dynamics of spinless particles, including the derivation of the ISCO. In Sect.~\ref{sec4}, we analyze the oscillatory motion, derive the fundamental frequencies, compute the QPO spectra using the ER3 and ER4 models, and investigate the collision of spinless particles. Sec.~\ref{sec:Vi} is devoted to studying the spin-particle dynamics using the Mathisson--Papapetrou-Dixon formalism. Finally, we summarize our findings and discuss their astrophysical implications in Sect.~\ref{sec6}.


\section{Einstein--Scalar Field--Massive Particle System \label{sec2}}

For the free relativistic particle with mass $m$, the action can be written as \cite{Landau1980}
\begin{align}
S = - \int m \, ds,
\end{align}

while in the presence of the external scalar field $\varphi$, it reads \cite{Breuer1973}
\begin{align}\label{eqgsint}
S = - \int m_* \, ds, \qquad m_* = m (1 + g_s \varphi),
\end{align}
Here $m_*$ is the effective mass of the test particle in the scalar field and $g_s$ is the dimensionless coupling parameter between the scalar field and massive particles. The action for a massive particle interacting with a scalar field is given by \cite{Misner1972,Breuer1973}:
\begin{align}
S[x^\mu, \varphi] = -\frac{1}{8\pi} \int d^4x \, \partial_\mu \varphi \partial^\mu \varphi 
- m \int ds \, (1 + g_s \varphi) \sqrt{-u^\mu u_\mu}.
\label{s}
\end{align}
It is important to note that the action describing a massive particle, as given in eq~\eqref{s}, remains applicable in curved spacetime. Accordingly, the general form of the action for the coupled Einstein–scalar field system, incorporating massive particles, can be extended in a natural way as follows.
\begin{align}
S[g_{\mu\nu}, x^\mu, \varphi] 
&= \frac{1}{16\pi} \int d^4x \sqrt{-g} 
\left( R - 2 \nabla_\mu \varphi \nabla^\mu \varphi \right) \nonumber \\
&\quad - m \int ds \, (1 + g_s \varphi)\, 
\sqrt{-u^\mu u_\mu}
\label{ss}
\end{align}
where $R$ is the Ricci scalar, $\nabla_\mu$ stands for covariant derivative from a scalar field, and $u^\mu = dx^\mu/ds$ is the four-velocity of the particle normalized as $u^\mu u_\mu = -1$.

Subsequently, by minimizing the action given in eq~\eqref{ss}, one obtains the equations of motion for the entire system, namely the Einstein field equations, the Klein–Gordon equation, and the geodesic equation.
\begin{align}
R_{\mu\nu} - \frac{1}{2} g_{\mu\nu} R 
&= 2 \nabla_\mu \varphi \nabla_\nu \varphi 
- g_{\mu\nu} \nabla_\alpha \varphi \nabla^\alpha \varphi \nonumber \\
&\quad + \frac{8\pi m}{\sqrt{-g}} 
\int ds \, (1 + g_s \varphi)\, 
\delta^{(4)}[x - z(s)]\, u_\mu u_\nu
\end{align}
\begin{align}
\nabla_\mu \nabla^\mu \varphi 
= \frac{4\pi m g_s}{\sqrt{-g}} \int ds \, \delta^{(4)}[x - z(s)],
\end{align}
\begin{align}
u^\nu \nabla_\nu u^\mu 
= \frac{g_s}{1 + g_s \varphi} (g^{\mu\nu} + u^\mu u^\nu) \nabla_\nu \varphi.
\label{um}
\end{align}
For the sake of simplicity, we consider the contribution of particle motion to the background spacetime and the scalar field to be negligible, implying that neither the spacetime geometry nor the scalar field configuration is affected. As a result, the equations of motion for the system reduce to
\begin{align}
R_{\mu\nu} &= 2 \nabla_\mu \varphi \nabla_\nu \varphi, \tag{8} \\
\nabla_\mu \nabla^\mu \varphi &= 0, \tag{9} \\
u^\nu \nabla_\nu u^\mu &= \frac{g_s}{1 + g_s \varphi} 
\left(g^{\mu\nu} + u^\mu u^\nu\right) \nabla_\nu \varphi. \tag{10}
\end{align}

\section{ Background Spacetime and particle dynamics \label{sec3}}

Following Ref.~\cite{Turimov2022}, we consider the generic three-parameter wormhole solution of the Einstein--scalar field equations,

\begin{align}
ds^2 = 
- f(r) dt^2
+ f(r)^{-1} dr^2
+f(r)^{-1} (r^2 + 2Mr + r_0^2)d\Omega^2
\end{align}
The corresponding metric function is given by
\begin{align}
    f(r)=\left(\frac{r + M + \sqrt{M^2 - r_0^2}}{r + M - \sqrt{M^2 - r_0^2}}\right)^{\frac{-\sigma}{\sqrt{M^2 - r_0^2}}}
\end{align}
The scalar-field configuration corresponding to the above solution is:
\begin{align}
\varphi(r) = \frac{1}{2}
\sqrt{1 - \frac{\sigma^2}{M^2 - r_0^2}}
\ln \left(
\frac{r + M + \sqrt{M^2 - r_0^2}}{r + M - \sqrt{M^2 - r_0^2}}
\right)
\label{fr}
\end{align}

Here, $M$ denotes the mass of the central object, $r_0$ characterizes the size of the wormhole throat, and $\sigma$ corresponds to the scalar charge parameter. The equation of motion for a massive particle, as given in eq~\eqref{um}, can be explicitly written in the background spacetime described by eq~\eqref{fr}.
\begin{widetext}
\begin{align}
\ddot t
+\frac{2}{\Delta(r)}
\left(
\sigma+
\frac{g_s\kappa\sqrt{1-\sigma^2/\kappa^2}}
{2+g_s\ln\mathcal R(r)\sqrt{1-\sigma^2/\kappa^2}}
\right)\dot r\,\dot t=0,
\end{align}

\begin{align}
\begin{aligned}
\ddot r + \frac{1}{\Delta(r)}\Bigg[
(-\sigma+\frac{2g_s\kappa(\sqrt{1-\sigma^2/\kappa^2}+\mathcal R(r)^{-\sigma/\kappa})}
{2+g_s\ln\mathcal R(r)\sqrt{1-\sigma^2/\kappa^2}})\dot r^{\,2} 
 + \sigma\,\mathcal R(r)^{-2\sigma/\kappa}\,\dot t^{\,2}
- (M-\sigma)\Delta(r)
\left(\dot\theta^{\,2}+\sin^2\theta\,\dot\varphi^{\,2}\right)
\Bigg] = 0,
\end{aligned}
\end{align}
\begin{align}
\ddot\theta
-\frac{2}{\Delta(r)}
\left(
-M+\sigma
-\frac{g_s\kappa\sqrt{1-\sigma^2/\kappa^2}}
{2+g_s\ln\mathcal R(r)\sqrt{1-\sigma^2/\kappa^2}}
\right)\dot r\,\dot\theta
-\sin\theta\cos\theta\,\dot\varphi^{\,2}
=0,
\end{align}
\begin{align}
\ddot\varphi +2\left[ \frac{1}{\Delta(r)} \left( M \sigma + \frac{g_s\kappa\sqrt{1-\sigma^2/\kappa^2}} {2+g_s\ln\mathcal R(r)\sqrt{1-\sigma^2/\kappa^2}} \right)\dot r +\cot\theta\,\dot\theta \right]\dot\varphi=0,
\end{align}
Here we use following notations
\begin{align}
    \Delta(r)=r^2+2Mr+r_0^2\ , \qquad \mathcal R(r)=\frac{r+M+\sqrt{M^2-r_0^2}}{r+M-\sqrt{M^2-r_0^2}}\ ,\qquad \kappa=\sqrt{M^2-r_0^2}\ .
\end{align}
\end{widetext}
Using \eqref{eqgsint} the Lagrangian for a massive particle in the presence of the scalar field can be expressed as
\begin{align}
    L=\frac{1}{2}(1+g_s\varphi)g_{\mu\nu}u^{\mu}u^{\nu}
\end{align}
and the four momentum of the massive particle is
\begin{align}
    P_{\mu}=\frac{\partial{L}}{\partial{u^{\mu}}}=(1+g_s\varphi)g_{\mu\nu}u^{\nu}
\end{align}
Since the spacetime does not explicitly depend on the coordinates $t$ and $\varphi$, the corresponding components of the four-momentum, $P_t$ and $P_\varphi$, are conserved. These conserved quantities are associated with the energy and angular momentum of the massive particle. Accordingly, the explicit forms of the specific energy $\mathcal{E}$ and specific angular momentum $\mathcal{L}$ can be written as follows.
\begin{align}
    \mathcal{E}&=-\frac{P_t}{m}=-(1+g_s\varphi)g_{tt}u^{t}\ ,\\
    \mathcal{L}&=(1+g_s\varphi)g_{\phi\phi}u^{\phi}
\end{align}
Using the normalization of the four-velocity of a particle, one can obtain
\begin{align}
g_{rr}\dot{r}^{2} + g_{\theta \theta} \dot{\theta}^{2} + 1 + \frac{1}{(1+g_s \varphi)^2} \left(\frac{\mathcal{E}^2}{g_{tt}} + \frac{\mathcal{L}^2}{g_{\phi \phi}} \right) = 0. 
\label{grrr}
\end{align}

Here, we focus on finding the innermost stable circular orbit (ISCO) position of a massive particle in the  spacetime in the presence of the scalar field. For the purpose of analyzing the qualitative features of particle motion, we restrict the trajectory of a massive particle to the equatorial plane, i.e., $\theta = \pi/2$ and $\dot{\theta} = 0$. Eq.~\eqref{grrr} can then be simplified to the following form after straightforward algebraic manipulations:
\begin{align}
\mathcal{E}^2 = (1+g_s \varphi)^2 \dot{r}^2 + U(r),
\end{align}
where
\begin{align}
U(r) = f (1+g_s \varphi)^2 + \frac{\mathcal{L}^2}{r^2+2Mr+r_0^2} f, \label{eff}
\end{align}
is the effective potential for a massive particle.
Using the conditions \[E^2 = U(r) \quad \text{and} \quad U'(r) = 0,\] one obtains the critical specific energy and angular momentum.

\begin{widetext}
\begin{align}
\mathcal{E} = - \frac{ \mathcal{R}^{-\sigma/\kappa} \, \big( 2 + B \ln \mathcal{R} \big) \, \big( 2 (M + r - \sigma - \kappa B) + (M + r - \sigma) B \ln \mathcal{R} \big) }{ 4 (M + r - 2 \sigma) }
\label{E}
\end{align}

\begin{align}
\mathcal{L} = \frac{ \Delta \, \mathcal{R}^{\sigma/\kappa} \, \big( 2 + B \ln \mathcal{R}\big) \, \big( 2 \sigma - 2 \kappa B + \sigma B \ln \mathcal{R} \big) }{ 4 (M + r - 2 \sigma) }
\label{L}
\end{align}
where
\[\kappa = \sqrt{M^2 - r_0^2}\ , \quad 
\mathcal{R} = \frac{M + r + \kappa}{M + r - \kappa}\ , \quad B = gs \sqrt{1 - \frac{\sigma^2}{\kappa^2}}\ ,\qquad \Delta = 2 M r + r^2 + r_0^2\ . \]

\end{widetext}
The ISCO is located at the stationary points of expressions \eqref{E} and \eqref{L}. Neglecting the interaction term (i.e., setting $g_s = 0$), the ISCO radius can be calculated as follows:
\begin{align}
  r_\text{ISCO}=1+3\sigma+\sqrt{-1+r_0^2+5\sigma^2}
\end{align}
 Fig.~\ref{isco} illustrates the dependence of the ISCO position of a massive particle on the wormhole throat radius $r_0$ for several values of the interaction parameter $g_s$. It can be seen that the ISCO radius increases monotonically with increasing throat size, indicating that larger wormhole throats tend to enlarge the region of stable circular motion. Moreover, increasing values of $g_s$ shift the ISCO curves upward, corresponding to larger stable orbit radii, whereas smaller or more negative values of $g_s$ lead to comparatively smaller ISCO radii. The influence of the interaction parameter becomes more pronounced for larger values of the throat radius, where the separation between different curves increases noticeably. This behavior demonstrates the combined effect of the scalarized wormhole geometry and interaction parameter on the effective potential governing particle motion.
\begin{figure}[ht!]
    \centering
    \includegraphics[width=0.9\linewidth]{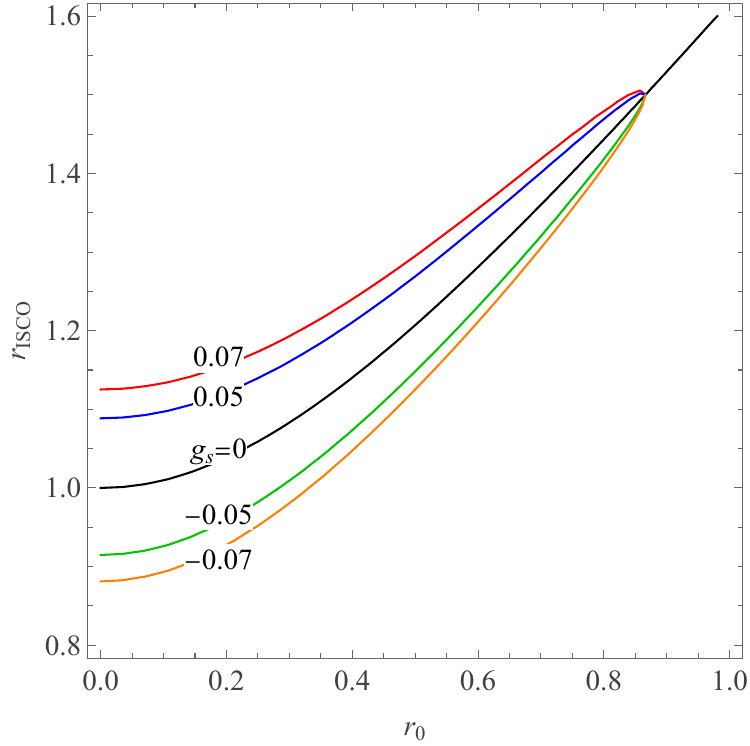}
    \caption{The position of the ISCO for a massive particle varies with $r_0$ and is influenced by the value of the coupling parameter.}
    \label{isco}
\end{figure}

\section{Oscillatory Motion of Test Particles \label{sec4}}

To analyze the oscillatory dynamics of neutral test particles, we consider small perturbations around stable circular orbits near the ISCO. When a particle is slightly displaced from its equilibrium position in the equatorial plane, it performs epicyclic motion described by linear harmonic oscillations.

\subsection{Fundamental Frequencies}

The investigation of particle motion around compact gravitational objects plays an important role in relativistic astrophysics. In general, the motion consists of the orbital (Keplerian) motion together with small radial and vertical oscillations. These oscillatory modes contain valuable information about the dynamical properties of the system and the underlying spacetime geometry near compact objects.
\begin{itemize}
\item \textbf{Orbital motion}. Orbital motion describes the azimuthal rotation of a test particle around a compact object along nearly circular geodesic trajectories. It is characterized by the Keplerian frequency, $\nu_{\phi}$, which specifies the angular rate of motion. This frequency defines the fundamental dynamical timescale of the system and plays an essential role in determining the structure and evolution of accretion disks.

\item \textbf{Radial oscillations}. Radial oscillations represent periodic displacements of a particle toward and away from the central object in the radial direction. These oscillations arise near stable circular orbits, and their behavior is governed by the geometry of spacetime and the physical parameters of the compact source.

\item \textbf{Vertical oscillations}. Vertical oscillations correspond to small deviations of a particle from the equatorial plane, leading to periodic motion above and below the plane while maintaining the underlying circular orbit.
\end{itemize}

The radial and vertical epicyclic oscillations are intrinsically connected to the orbital motion characterized by the Keplerian frequency. Together, these quantities form the set of fundamental frequencies of particle motion. 

In this work, we focus on the oscillatory motion of a massive particle near the stable circular orbit around  three parametric scalarized wormhole. In this case, the four-velocity of the particle can be written as
where $\Omega = \frac{d\phi}{dt}$ represents the angular velocity of the particle \cite{Turimov2024JNW}.
\begin{align}
u^\mu = \dot{t}\,(1,0,0,\Omega),
\end{align}
\begin{align}
    \Omega^2=-\frac{(1+g_s\varphi)\partial_rg_{tt}-2g_sg_{tt}\varphi'(r)}{(1+g_s\varphi)\partial_rg_{\phi\phi}-2g_sg_{\phi\phi}\varphi'(r)}
\end{align}

Using the spacetime described by eq~(\ref{fr}), the Keplerian frequency can be obtained as follows:
\begin{align}
\Omega=\sqrt{\frac{\mathcal{R}^{\frac{-2\sigma}{\kappa}}(2\sigma+\kappa B)+\sigma B\ln\mathcal{R}}{\Delta(2(M+r-\sigma+\kappa B)+(M+r-\sigma)B\ln\mathcal{R})}}
\end{align}
%


%
From the normalization condition of the four-velocity for a massive particle, one obtains
\begin{align}
g_{rr} \dot{r}^2 + g_{\theta\theta} \dot{\theta}^2 + V(r,\theta) = 0,
\label{um}
\end{align}
where $V(r,\theta)$ is the effective potential explicitly given in eq \eqref{eff}. After performing the following transformations $r \rightarrow r_0 + \delta r$ and $\theta \rightarrow \theta_0 + \delta \theta$ in equation (34), 2D-harmonic oscillator can be derived as follows:

\begin{align}
\begin{aligned}
&g_{rr}(\dot{r}_0+\dot{\delta r})^2 +g_{\theta\theta}(\dot{\theta}_0+\dot{\delta\theta})^2+V(r_0,\theta_0) \\
&+\partial_r V(r_0,\theta_0)\delta r+\partial_\theta V(r_0,\theta_0)\delta\theta+\partial_r\partial_\theta V(r_0,\theta_0)\delta r\,\delta\theta \\&+\frac{1}{2}\partial_r^2 V(r_0,\theta_0)\delta r^2+\frac{1}{2}\partial_\theta^2 V(r_0,\theta_0)\delta\theta^2+\cdots = 0 .
\end{aligned}\label{111}
\end{align}

The points $(r_0,\theta_0)$ correspond to the stationary points of the effective potential function $V(r,\theta)$. Expanding the equations of motion around these equilibrium positions and considering small perturbations in the radial and vertical directions, namely $\delta r$ and $\delta\theta$, one obtains the equations governing the oscillatory motion of the particle:

\begin{align}
g_{rr}\left(\frac{d^2}{dt^2}+\Omega_r^2\right)\delta r+g_{\theta\theta}\left(\frac{d^2}{dt^2}+\Omega_\theta^2\right)\delta\theta=0,
\end{align}\label{222}
where the radial epicyclic frequency $\Omega_r$ is defined by

\begin{align}
\Omega_r^2=\frac{1}{2g_{rr}\dot{t}^{\,2}}\frac{\partial^2 V}{\partial r^2}=\frac{-g_{tt}-\Omega^2 g_{\phi\phi}}{2g_{rr}}\frac{\partial^2 V}{\partial r^2},
\end{align}\label{rad}
while the vertical epicyclic frequency $\Omega_\theta$ takes the form

\begin{align}
\Omega_\theta^2=\frac{1}{2g_{\theta\theta}\dot{t}^{\,2}}\frac{\partial^2 V}{\partial \theta^2}=\frac{-g_{tt}-\Omega^2 g_{\phi\phi}}{2g_{\theta\theta}}\frac{\partial^2 V}{\partial \theta^2}.
\end{align}\label{kep}
The frequencies can be expressed in Hertz (Hz) using the following equation:
\begin{align}\label{nu_rel}
\nu_{i}=\frac{1}{2\pi}\frac{c^{3}}{GM} \, \Omega_{i}[{\rm Hz}],
\end{align}
where $c = 3 \times 10^{10} \, \text{cm s}^{-1}$ is the speed of light in a vacuum and $G = 6.67 \times 10^{-8} \, \text{cm}^3 \text{g}^{-1} \text{s}^{-2}$ is the Newtonian gravitational constant.

\begin{figure*}\centering
\includegraphics[width=0.3\linewidth]{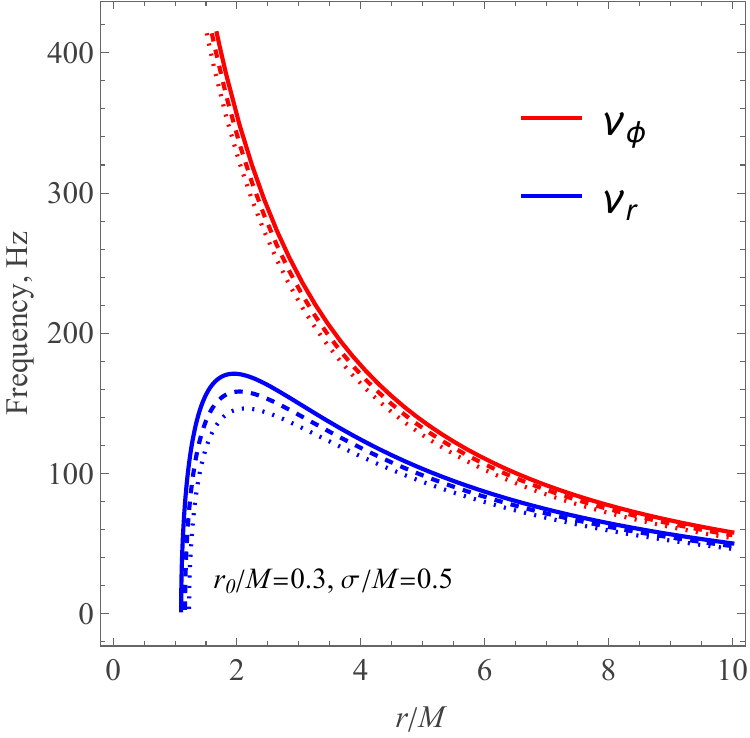}
\includegraphics[width=0.3\linewidth]{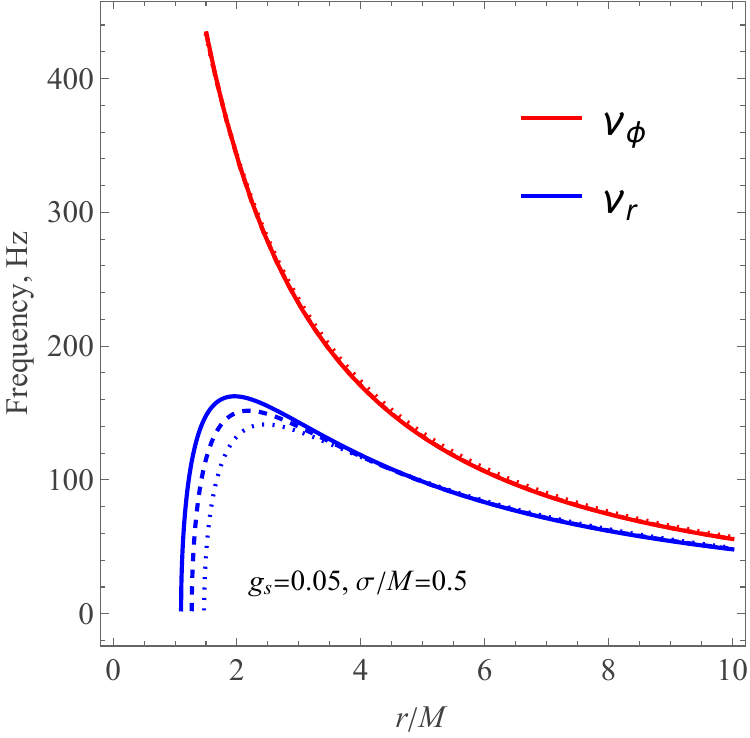}
\includegraphics[width=0.3\linewidth]{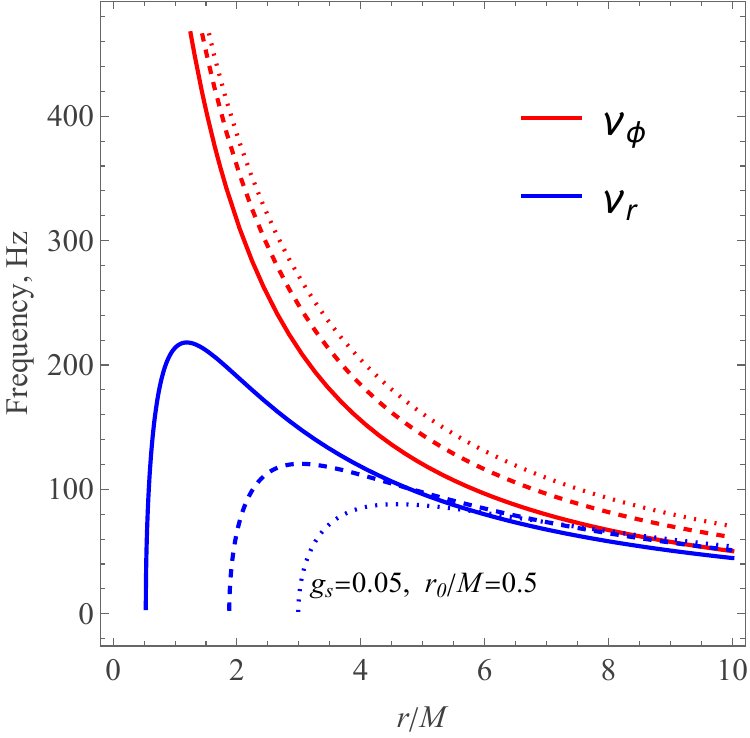}

\caption{Radial dependence of the azimuthal ($\nu_{\phi}$) and radial ($\nu_{r}$) frequencies of massive particles in the three-parameter scalarized wormhole spacetime. In the left panel, the solid, dashed, and dotted curves correspond to $g_s=0.01$, $0.05$, and $0.09$, respectively. In the middle panel, they represent $r_0/M=0.1$, $0.5$, and $0.8$, while in the right panel they correspond to $\sigma/M=0.2$, $0.4$, and $0.6$.}
\label{fun}
\end{figure*}

The radial dependence of the fundamental frequencies is shown in Fig.~\ref{fun}. The red curves represent the orbital frequency $\nu_{\phi}$, whereas the blue curves correspond to the radial epicyclic frequency $\nu_{r}$. The azimuthal frequency decreases monotonically with increasing radial distance, while the radial epicyclic frequency initially increases, reaches a maximum, and then gradually decreases. Throughout the considered radial range, $\nu_{\phi}$ remains larger than $\nu_{r}$.

The left panel illustrates the effect of the scalar--particle coupling parameter $g_s$. Increasing $g_s$, from the solid to the dashed and dotted curves, enhances both the azimuthal and radial epicyclic frequencies. The middle panel shows the influence of the wormhole throat radius $r_0/M$. As $r_0/M$ increases, both characteristic frequencies decrease, with the effect being more pronounced for the radial epicyclic frequency. The right panel demonstrates the influence of the scalar charge parameter $\sigma/M$. Increasing $\sigma/M$ enhances the azimuthal frequency $\nu_{\phi}$, whereas it suppresses the radial epicyclic frequency $\nu_{r}$ and shifts its maximum toward larger radial distances.

\subsection{Quasiperiodic oscillation}

In the vicinity of compact astrophysical objects, particle motion deviates from purely circular trajectories due to the presence of small radial and vertical perturbations around the equilibrium orbit. These oscillatory motions, superimposed on the orbital rotation, generate complex trajectories and are believed to play a crucial role in the origin of quasi-periodic oscillations (QPOs) observed in X-ray binary systems. Such QPOs are commonly associated with oscillatory processes occurring in accretion disks surrounding black holes and neutron stars.

These oscillations have an important impact on the dynamical behavior and stability of accretion disks. They influence both the spectral properties and the temporal variability of the emitted radiation. Consequently, studying the characteristic frequencies of these oscillations provides essential information about the physical parameters of compact objects, including their mass, rotation, and the underlying spacetime geometry.

Although wormholes themselves do not emit electromagnetic radiation, their intense gravitational fields strongly affect the motion of nearby matter and significantly curve the surrounding spacetime. Matter captured by the gravitational field forms an accretion disk composed of gas and plasma orbiting the compact object. As this matter gradually spirals inward, dissipative processes heat the disk to extremely high temperatures, leading to powerful electromagnetic emission over a broad range of frequencies.

\textbf{Epicyclic Resonance Models.}
The epicyclic resonance (ER) model explains quasi-periodic oscillations as a result of nonlinear resonant interactions between oscillation modes in accretion disks around compact objects. In this framework, the oscillation frequencies are directly related to the orbital and epicyclic frequencies of particles moving along circular geodesics.

\begin{figure*}\centering
\includegraphics[width=0.3\linewidth]{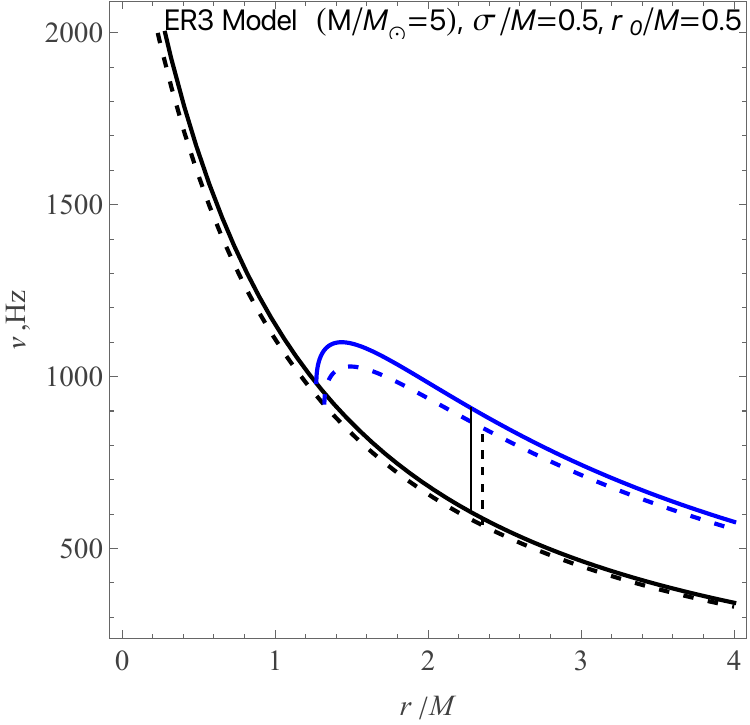}
\includegraphics[width=0.3\linewidth]{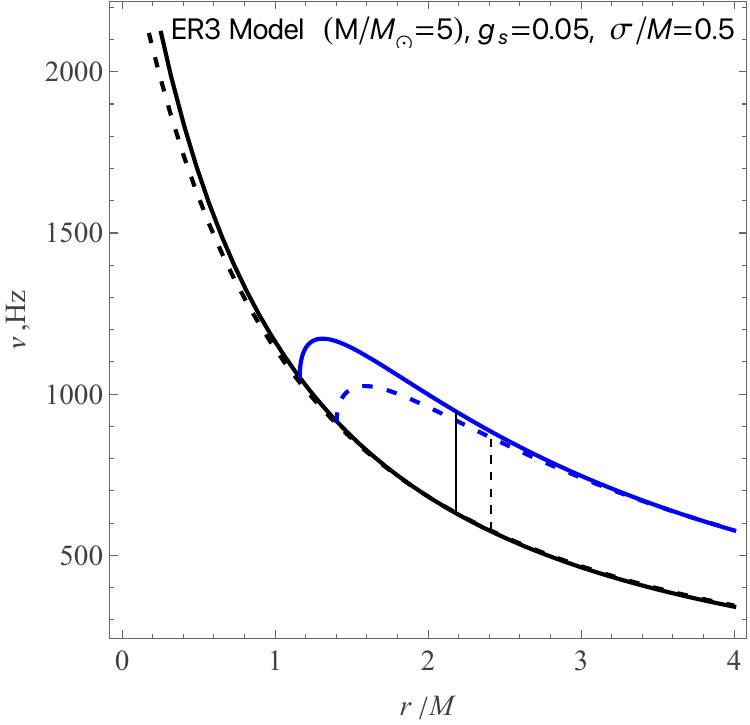}
\includegraphics[width=0.3\linewidth]{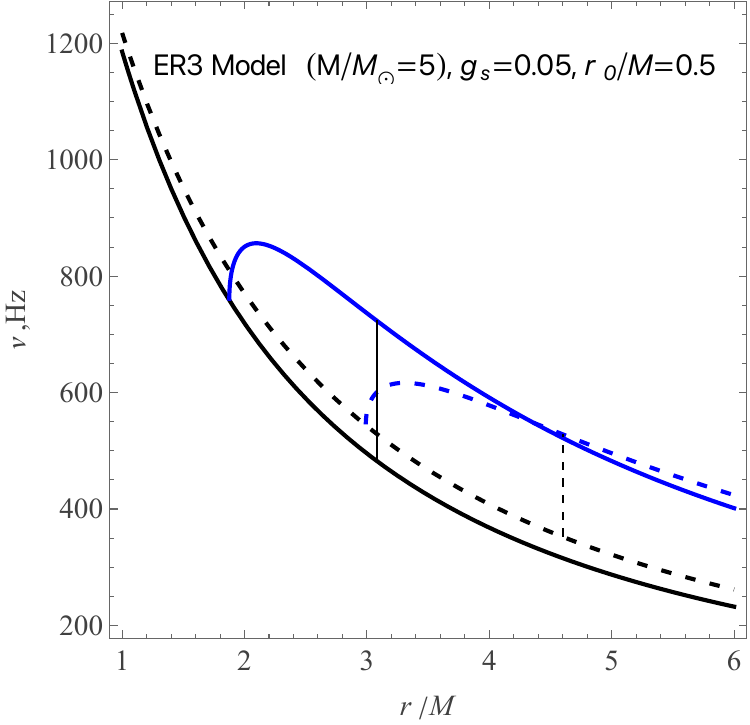}

\caption{Radial dependence of the upper ($\nu_{\rm U}$) and lower ($\nu_{\rm L}$) twin-peak QPO frequencies in the ER3 model. In the left panel, the solid and dashed curves correspond to $g_s=0.05$ and $0.09$, respectively. In the middle panel, they correspond to $r_0/M=0.3$ and $0.7$, respectively, while in the right panel they represent $\sigma/M=0.6$ and $0.8$, respectively. The blue and black curves denote the upper ($\nu_{\rm U}$) and lower ($\nu_{\rm L}$) frequencies, respectively.}
\label{ratio1}
\end{figure*}

\begin{figure*}\centering
\includegraphics[width=0.3\linewidth]{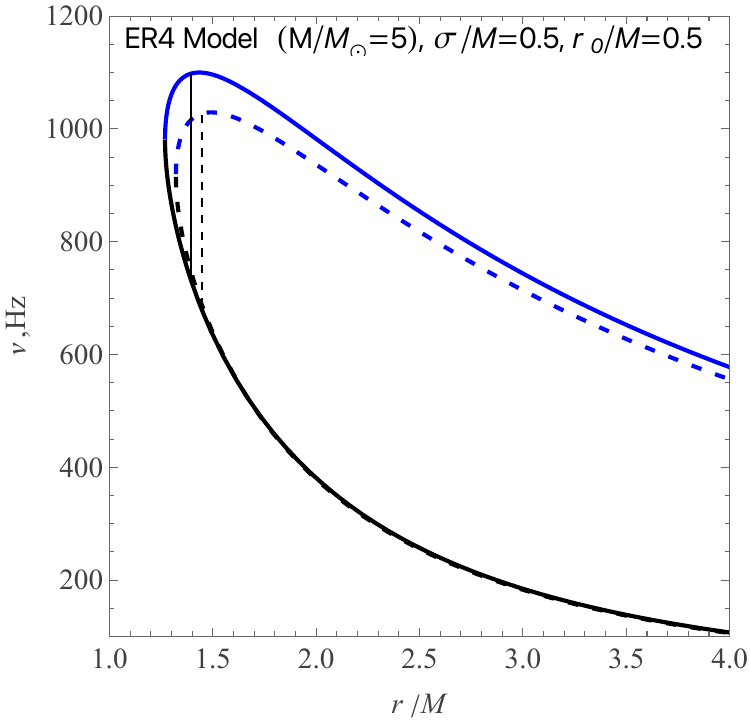}
\includegraphics[width=0.3\linewidth]{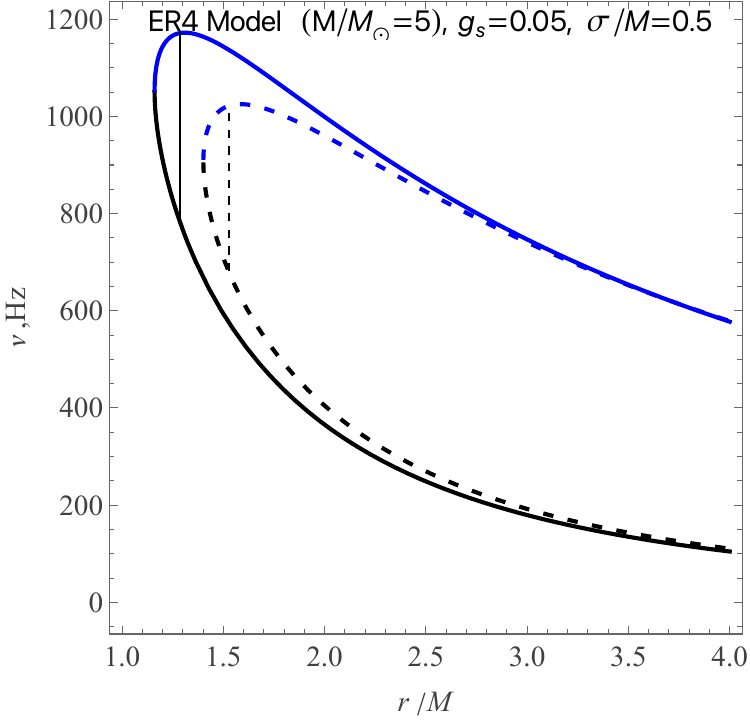}
\includegraphics[width=0.3\linewidth]{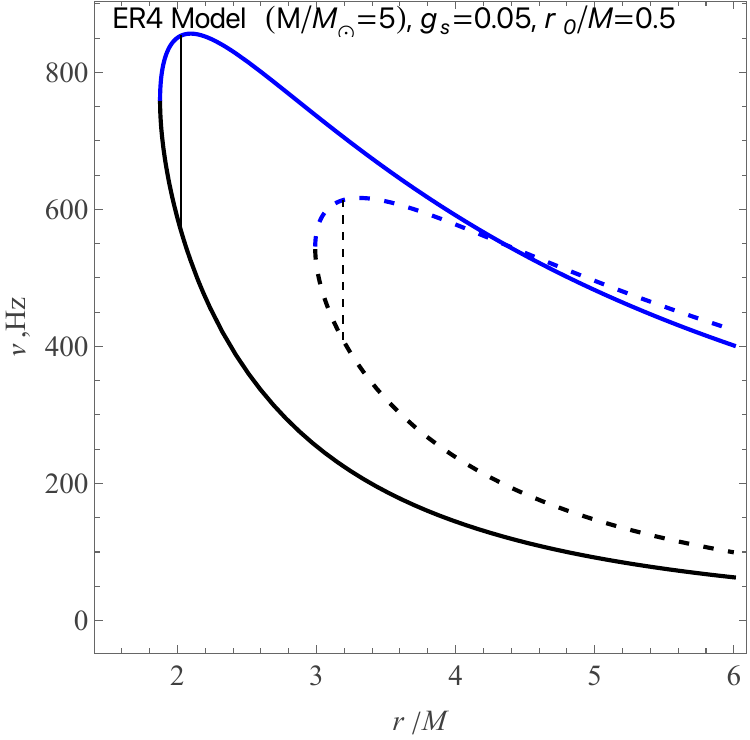}

\caption{Radial dependence of the upper ($\nu_{\rm U}$) and lower ($\nu_{\rm L}$) twin-peak QPO frequencies in the ER4 model. In the left panel, the solid and dashed curves correspond to $g_s=0.05$ and $0.09$, respectively. In the middle panel, they correspond to $r_0/M=0.3$ and $0.7$, respectively, while in the right panel they represent $\sigma/M=0.6$ and $0.8$, respectively. The blue and black curves denote the upper ($\nu_{\rm U}$) and lower ($\nu_{\rm L}$) frequencies, respectively. The vertical dashed lines mark the resonance radii satisfying $\nu_{\rm U}/\nu_{\rm L}=3:2$.}
\label{ratio2}
\end{figure*}

Different versions of the ER model arise from distinct combinations of the fundamental frequencies. In the present work, we consider the ER3 and ER4 models, which are constructed from the radial and vertical epicyclic frequencies, denoted by $\nu_r$ and $\nu_\theta$, respectively.
For the ER3 model, the upper and lower QPO frequencies are expressed as
\begin{align}
\nu_U = \nu_\theta + \nu_r, \qquad
\nu_L = \nu_\theta ,
\end{align}
whereas for the ER4 model they are given by
\begin{align}
\nu_U = \nu_\theta + \nu_r, \qquad
\nu_L = \nu_\theta - \nu_r .
\end{align}
Figs. ~\ref{ratio1} and~\ref{ratio2} illustrate the radial dependence of the upper and lower twin-peak QPO frequencies in the ER3 and ER4 models, respectively. In both models, the radius satisfying the characteristic $3:2$ frequency ratio shifts toward larger values as $g_s$, $r_0/M$, and $\sigma/M$ increase. This result indicates that the resonance region moves farther from the wormhole throat as the scalar coupling and the wormhole parameters become larger.

\begin{figure*}\centering
\includegraphics[width=0.25\linewidth]{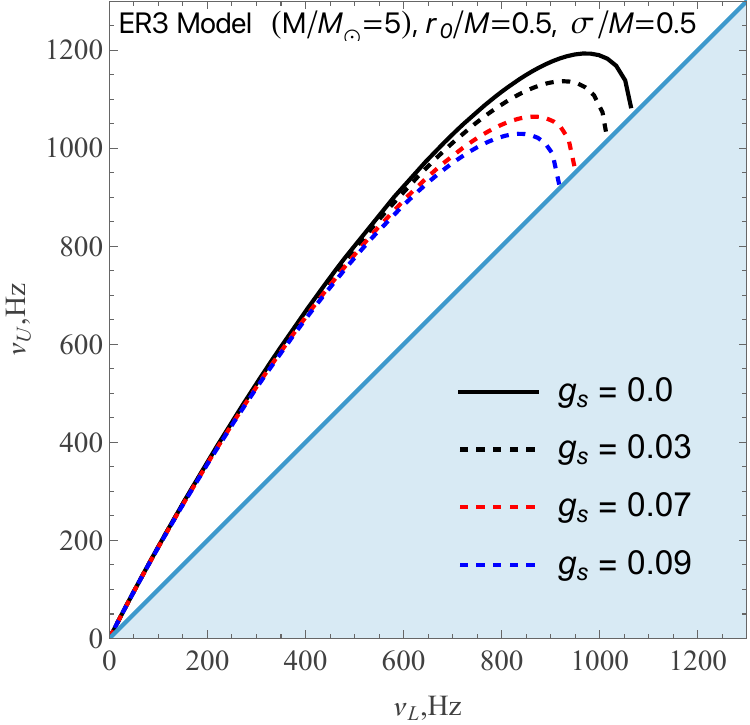}
\includegraphics[width=0.25\linewidth]{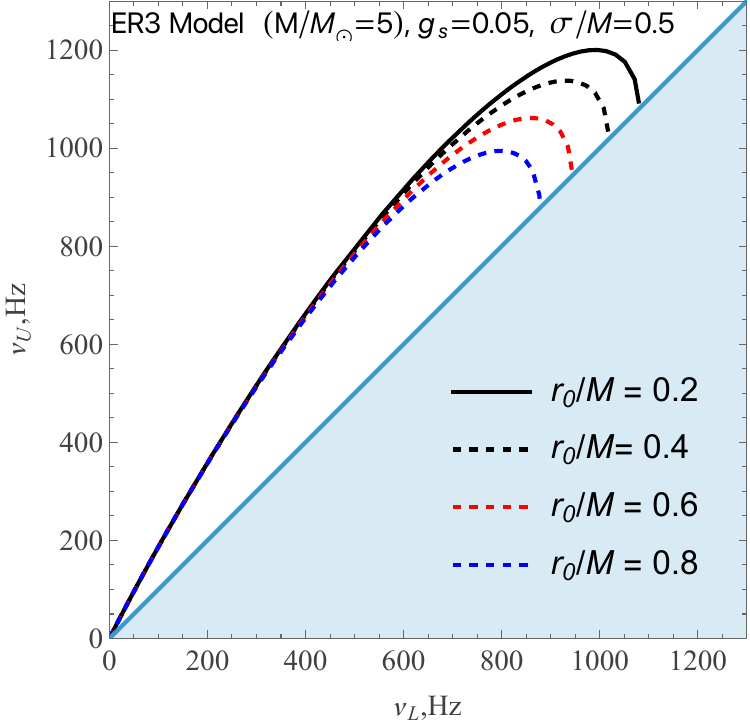}
\includegraphics[width=0.25\linewidth]{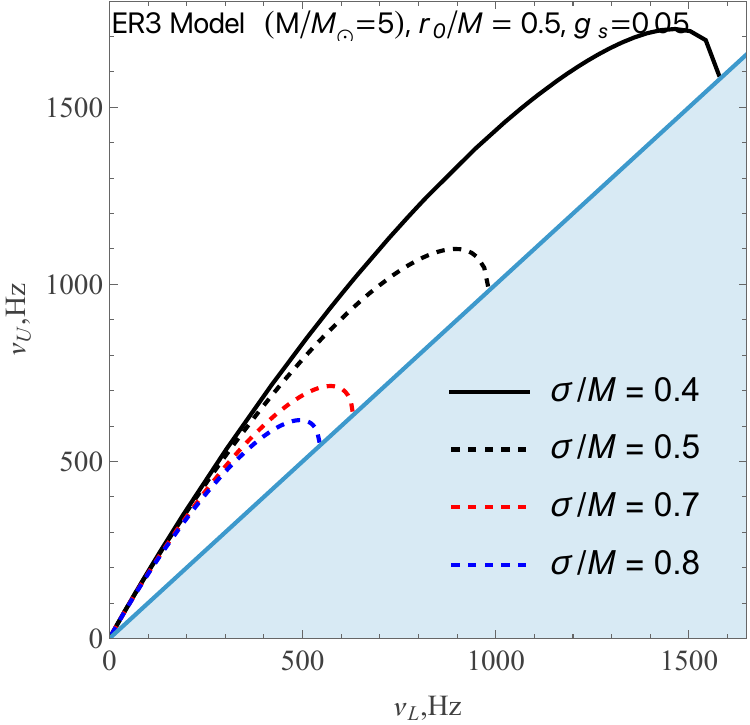}

\caption{Relationship between upper and lower frequencies of the twin peak QPO in the ER3 model with mass $M=5M_{\odot}$}. \label{ul3}
 \end{figure*}

\begin{figure*}\centering
\includegraphics[width=0.25\linewidth]{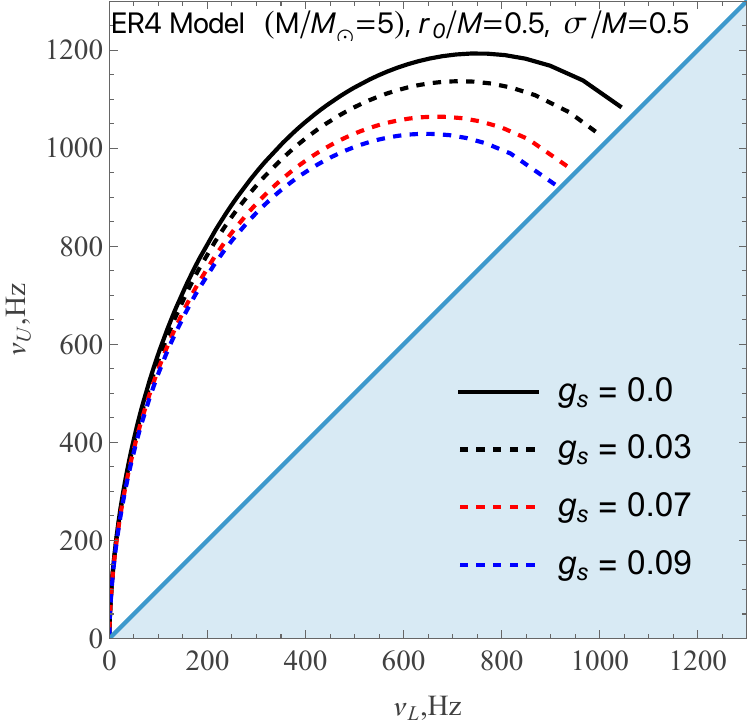}
\includegraphics[width=0.25\linewidth]{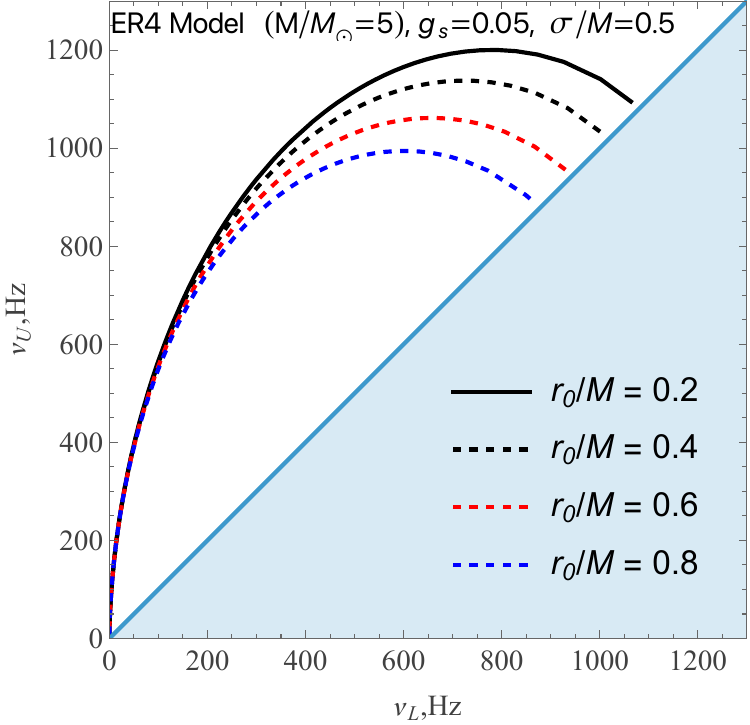}
\includegraphics[width=0.25\linewidth]{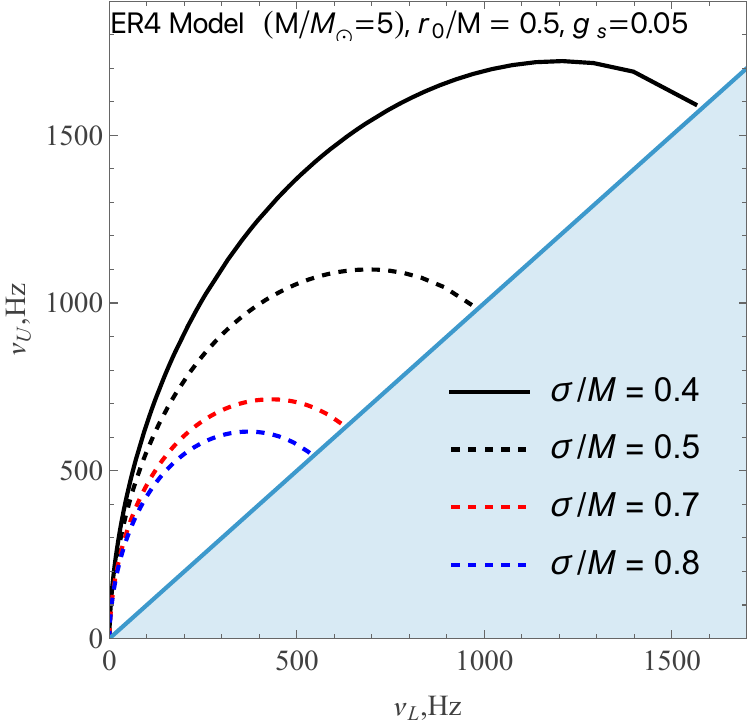}

\caption{Relationship between upper and lower frequencies of the twin peak QPO in the ER4 model with mass $M=5M_{\odot}$}. \label{ul4}
 \end{figure*}

Figs~\ref{ul3} and~\ref{ul4} display the correlation between the upper ($\nu_{\rm U}$) and lower ($\nu_{\rm L}$) QPO frequencies for different values of $g_s$, $r_0/M$, and $\sigma/M$. The calculated frequency correlations are sensitive to both the scalar--particle interaction and the geometrical parameters of the wormhole. The shaded regions correspond to radii below the ISCO, where stable circular motion is not permitted. Consequently, the epicyclic-resonance interpretation adopted here is physically applicable only outside the ISCO.

 
\begin{figure*}\centering
\includegraphics[width=0.3\linewidth]{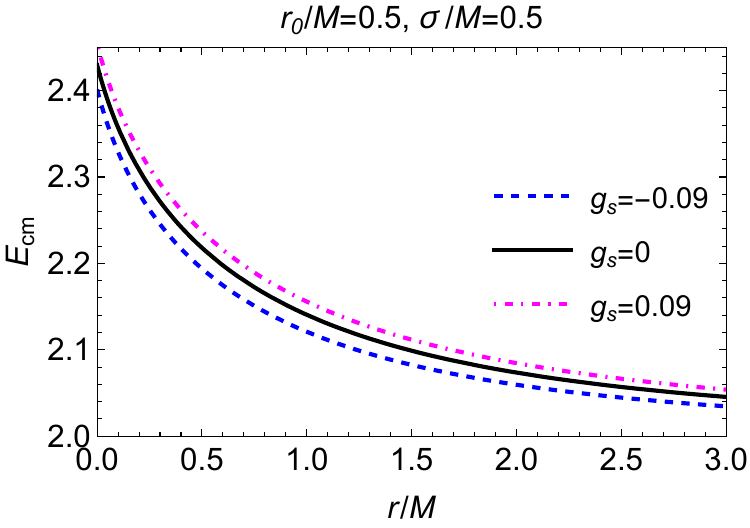}
\includegraphics[width=0.3\linewidth]{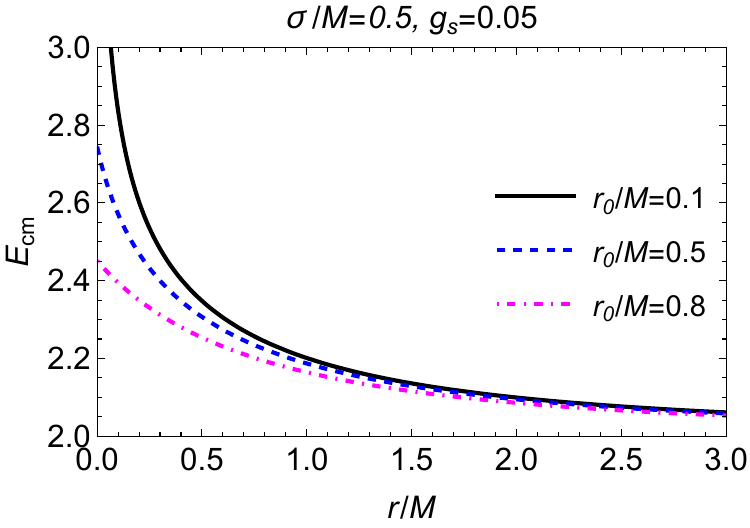}
\includegraphics[width=0.3\linewidth]{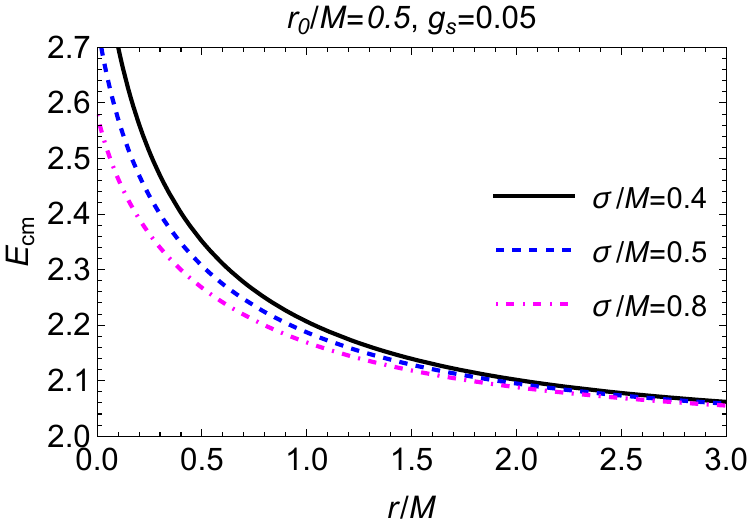}

\caption{Radial dependence of the dimensionless center-of-mass energy $\mathcal{E}_{\rm cm}$ for different values of the scalar coupling parameter $g_s$ (left), throat radius $r_0/M$ (middle), and scalar field parameter $\sigma/M$ (right).}
\label{ecm1}
\end{figure*}

\subsection{Collision of particles}

This section investigates the collision of two neutral spinless test particles with identical rest mass $\mu$. The particles move in the equatorial plane with different four-velocities. The center-of-mass (CM) energy of the colliding particles is given by
\begin{align}
\frac{E_{\rm cm}}{\sqrt{2}m_0}
=
\sqrt{1-g_{\alpha\beta}u_1^{\alpha}u_2^{\beta}},
\end{align}
where the four-velocity of each particle is

\begin{align}
u_i^{\alpha}=\left(\frac{\mathcal{E}_i}{(1+g_s\varphi)g_{tt}},-Y_i,0,\frac{\mathcal{L}_i}{(1+g_s\varphi)g_{\phi\phi}}\right),
\end{align}
with

\begin{align}
Y_i=\sqrt{\frac{-1-g_{tt}(u_i^t)^2-g_{\phi\phi}(u_i^\phi)^2}{g_{rr}}}\ ,
\end{align}
or equivalently,

\begin{align}
Y_i=\sqrt{-\mathcal{R}^{-\sigma/\kappa}-\frac{4\mathcal{L}_i^2\mathcal{R}^{-2\sigma/\kappa}}{\Delta(2+B\ln\mathcal{R})^2}+\frac{4\mathcal{E}_i^2\mathcal{R}^{2\sigma/\kappa}}{2+B\ln\mathcal{R}}}.
\end{align}

Here, $\mathcal{E}_i$ and $\mathcal{L}_i$ ($i=1,2$) denote the conserved energy and angular momentum per unit mass of the $i$th particle, respectively. Consequently, the center-of-mass energy takes the form
\begin{widetext}
\begin{align}
\frac{E_{\rm cm}}{\sqrt{2}m(1+g_s \varphi)} =\sqrt{1-\frac{\mathcal{E}_1\mathcal{E}_2}{(1+g_s\varphi)^2g_{tt}}-g_{rr}Y_1Y_2-\frac{\mathcal{L}_1\mathcal{L}_2}{(1+g_s\varphi)^2g_{\phi\phi}}}.
\end{align}
\end{widetext}
Fig. ~\ref{ecm1} illustrates the radial dependence of the dimensionless center-of-mass energy $\mathcal{E}_{\rm cm}$ for collisions of two  particles in the scalarized wormhole spacetime. In all panels, the collision energy reaches its maximum in the vicinity of the wormhole throat and decreases monotonically with increasing radial distance. This behavior reflects the stronger gravitational field near the throat, where particles acquire higher relative velocities before collision.

The left panel shows the influence of the scalar coupling parameter $g_s$. Increasing $g_s$ systematically enhances the center-of-mass energy over the entire radial range, indicating that the interaction between the particles and the background scalar field increases the available collision energy.
The middle panel demonstrates the effect of the throat radius $r_0/M$. For fixed values of the scalar coupling and scalar charge, smaller throat radii produce noticeably larger collision energies, whereas increasing $r_0/M$ suppresses the collision energy. This indicates that a more compact wormhole geometry provides a deeper gravitational potential, leading to more energetic particle collisions near the throat.

The right panel presents the dependence on the scalar charge parameter $\sigma/M$. It is evident that increasing $\sigma/M$ also reduces the center-of-mass energy throughout the considered radial region. Consequently, both a larger throat radius and a stronger scalar charge weaken the acceleration of infalling particles, whereas the scalar coupling parameter acts in the opposite direction by enhancing the collision energy.

\section{Spinning particle motion \label{sec:Vi}}
To investigate the dynamics of spinning particles in the scalarized wormhole spacetime, we employ the Mathisson--Papapetrou--Dixon (MPD) formalism together with the Tulczyjew spin supplementary condition. The spin--curvature interaction leads to deviations from geodesic motion and modifies the conserved quantities of the particle. In this section, we derive the equations of motion and the corresponding effective potential, which forms the basis for the analysis of circular orbits and particle collisions presented in the following sections.

\subsection{Equations of motion}

The motion of spinning test particles is governed by theMathisson--Papapetrou--Dixon (MPD) equations, which describe the interaction between the intrinsic spin of the particle and the spacetime curvature. For a spinning particle non-minimally coupled to a scalar field, the complete equations of motion generally differ from the standard MPD equations owing to the presence of an additional force proportional to the scalar-field gradient. In the present work, however, we restrict ourselves to the weak-coupling regime ($|g_s|\ll1$), where this contribution can be neglected as a higher-order correction. Consequently, the scalar field is incorporated only through the effective mass $m^*=m(1+g_s\varphi)$, while the particle dynamics are described by the standard MPD equations, given by \cite{Mathisson1937,Papapetrou1951}
\begin{align}
\frac{Dp^\alpha}{d\lambda}&=-\frac12R^\alpha_{\ \beta\mu\nu}u^\beta S^{\mu\nu},\\
\frac{DS^{\alpha\beta}}{d\lambda}&=p^\alpha u^\beta-p^\beta u^\alpha .
\label{MPD}
\end{align}
Here, $D/d\lambda\equiv u^\gamma\nabla_\gamma$ denotes the covariant derivative along the particle worldline, while $R^\alpha_{\ \beta\mu\nu}$, $p^\alpha$, $u^\alpha$, and $S^{\alpha\beta}$ denote the Riemann curvature tensor, four-momentum, four-velocity, and antisymmetric spin tensor, respectively.

Since
\begin{align}
S^{\alpha\beta}=-S^{\beta\alpha},
\end{align}
the spin tensor has six independent components.

The spin-curvature interaction $R^\alpha_{\ \beta\mu\nu}u^\beta S^{\mu\nu}$ causes deviations from geodesic motion. For $s=0$, the MPD equations reduce to the geodesic equations.

The MPD equations are supplemented by the Tulczyjew spin supplementary condition,
\begin{align}
S^{\mu\nu}p_{\nu}=0,
\label{tul}
\end{align}
which removes the remaining degrees of freedom associated with the spin tensor.\cite{Tulczyjew1959}
The particle mass and spin magnitude are conserved,
\begin{align}
p^\mu p_\mu&=-m^2(1+g_s\varphi)^2,
\label{pnorm}\\
S^{\mu\nu}S_{\mu\nu}&=2S^2.
\label{snorm}
\end{align}

The conserved quantities associated with spacetime symmetries are obtained by introducing the corresponding Killing vectors $\xi^\alpha$. In the MPD formalism, the conserved quantity takes the form

\begin{align}
C_{\xi}=p^\alpha\xi_\alpha-\frac12S^{\alpha\beta}\nabla_\beta\xi_\alpha .
\end{align}

Because the spin tensor $S^{\alpha\beta}$ is antisymmetric, whereas the Christoffel symbols $\Gamma^{\nu}_{\alpha\beta}$ are symmetric in their lower indices, their contraction identically vanishes,
\begin{align}
S^{\alpha\beta}\Gamma^{\nu}_{\alpha\beta}=0.
\end{align}
As a result, the covariant derivative acting on the Killing vector reduces to an ordinary derivative, and the conserved quantity can be expressed as

\begin{align}
C_{\xi}=p^\alpha\xi_\alpha-\frac12S^{\alpha\beta}\partial_\beta\xi_\alpha .
\end{align}

For a static, spherically symmetric spacetime, the metric is written as

\begin{align}
ds^2=g_{tt}dt^2+g_{rr}dr^2+g_{\theta\theta}d\theta^2+g_{\phi\phi}d\varphi^2,
\end{align}
where all metric functions depend only on the radial coordinate $r$. Owing to the stationarity and spherical symmetry of the spacetime, two independent Killing vectors exist, corresponding to time translations and rotations,

\begin{align}
\xi^\alpha_{(t)}=\delta^\alpha_t,\qquad\xi^\alpha_{(\phi)}=\delta^\alpha_\phi.
\end{align}

These symmetries give rise to two conserved quantities, namely the particle energy and its total angular momentum $J$, defined as $J = L + S$, with $L$ and $S$ representing the orbital and spin angular momentum, respectively.\cite{Abdulkhamidov2023}

\begin{align}
E&=-p_t+\frac12g_{tt,r}S^{tr},\\
J&=p_\varphi-\frac12 g_{\phi\phi,r}S^{\phi r}.
\end{align}

Since the present analysis is restricted to equatorial motion $(\theta=\pi/2)$, the polar component of the four-momentum vanishes, $p_\theta=0$. Consequently, only three independent components of the spin tensor remain nonzero, satisfying

\begin{align}
S^{\theta\alpha}=0.
\end{align}
Applying the Tulczyjew spin supplementary condition eq~\eqref{tul} to the remaining non-vanishing components of the spin tensor leads to

\begin{align}
S^{t\phi}=-\frac{p_r}{p_\phi}S^{tr},\qquad S^{r\phi}=\frac{p_t}{p_\phi}S^{tr}.
\label{sp}
\end{align}

Using the four-momentum normalization condition eq~\eqref{pnorm}, the radial component of the momentum can be written as
\begin{align}
p_r^2=g_{rr}\left[-g^{tt}p_t^2-g^{\varphi\varphi}p_\phi^2-m^2(1+g_s\varphi)^2\right].
\label{eq:pr}
\end{align}
Combining Eqs.~(\ref{sp}) and(\ref{eq:pr}) together with

\begin{align}
    S^{\alpha\beta}S_{\alpha\beta}=2m^2(1+g_s\varphi)^2s^2,
\end{align}
yields

\begin{align}
S^{tr}=-\frac{p_\phi s}{\sqrt{g_{tt}g_{rr}g_{\phi\phi}}},\qquad S^{\phi r}=\frac{p_t s}{\sqrt{-g_{tt}g_{rr}g_{\phi\phi}}}.
\label{eq:Str}
\end{align}

The covariant momentum components are given by
\begin{align}
p_t &=\frac{-E+s\mathcal{A}J}
     {1-s^{2}\mathcal{C}},\\
p_{\phi} &=\frac{J+s\mathcal{B}E}     {1-s^{2}\mathcal{C}}.
\end{align}
where

\begin{align}
\mathcal{A}&=\frac{g_{tt,r}}{2\sqrt{-g_{tt}g_{rr}g_{\phi\phi}}},\\
\mathcal{B}&=\frac{g_{\phi\phi,r}}{2\sqrt{-g_{tt}g_{rr}g_{\phi\phi}}},\\
\mathcal{C}&=\frac{g_{tt,r}g_{\phi\phi,r}}{4g_{rr}g_{tt}g_{\phi\phi}}.
\end{align}

Finally, using $p^\mu=g^{\mu\nu}p_\nu$, the radial equation becomes
\begin{align}
p_r^2=\frac{\alpha E^2+\delta E+\gamma}{\rho}.
\label{prad}
\end{align}
\subsection{Effective potential}

The effective potential is obtained from the radial equation by imposing the turning-point condition
\begin{align}
p_r=0,
\end{align}
which yields a quadratic equation for the conserved energy $E$. Accordingly, the radial equation  \eqref{prad} can be factorized as

\begin{align}
p_r^2=\frac{\alpha}{\rho}(E-V_{\rm eff}^{+})(E-V_{\rm eff}^{-}),
\label{eq:factor}
\end{align}
where the two branches of the effective potential are

\begin{align}
V_{\rm eff}^{\pm}=\frac{-\delta\pm\sqrt{\delta^24\alpha\gamma}}{2\alpha}.
\label{eq:Veff}
\end{align}

The coefficients are given by

\begin{align}
\begin{aligned}
\alpha&=4g_{tt}\left(-4g_{\phi\phi}^{2}g_{rr}+s^{2}(g'_{\phi\phi})^{2}\right),\\
\delta&=16Js\sqrt{-g_{\phi\phi}g_{rr}g_{tt}}\left(-g_{tt}g'_{\phi\phi}+g_{\phi\phi}g'_{tt}\right),\\
\rho&=\left(-4g_{\phi\phi}g_{rr}g_{tt}+s^{2}g'_{\phi\phi}g'_{tt}\right)^2,\\
\gamma&=-\frac{\left(-4g_{\phi\phi}g_{rr}g_{tt}+s^{2}g'_{\phi\phi}g'_{tt}\right)^2[m(1+g_s\phi)]^2}{g_{rr}}\\&\qquad-16J^{2}g_{\phi\phi}g_{rr}g_{tt}^{}+4J^{2}s^{2}g_{\phi\phi}(g'_{tt})^{2}.
\end{aligned}
\label{eq:coefficients}
\end{align}
Introducing the dimensionless variables

\begin{align}
\mathcal{E}=\frac{E}{m},\qquad \mathcal{L}=\frac{L}{m}, \qquad
s=\frac{S}{mM},
\end{align}
the allowed energy regions are

\begin{align}
\mathcal{E}&\le\mathcal{V}_{\rm eff}^{-},\\
\mathcal{E}&\ge\mathcal{V}_{\rm eff}^{+}.
\end{align}

\begin{figure*}\centering
\includegraphics[width=0.40\linewidth]{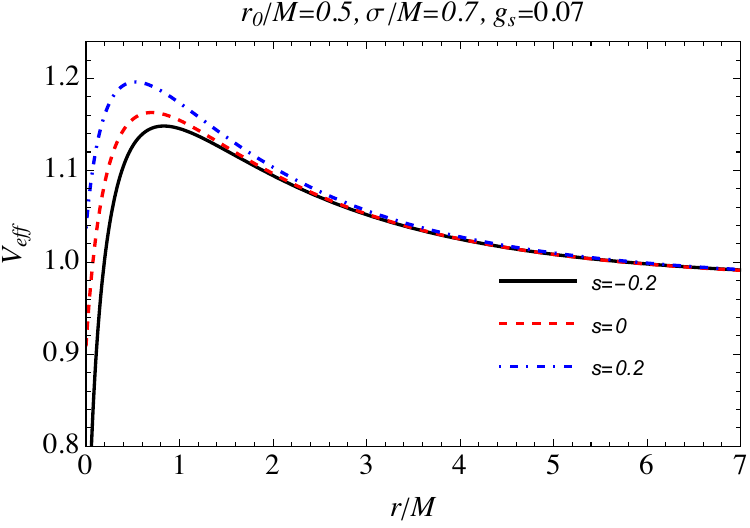}
\includegraphics[width=0.40\linewidth]{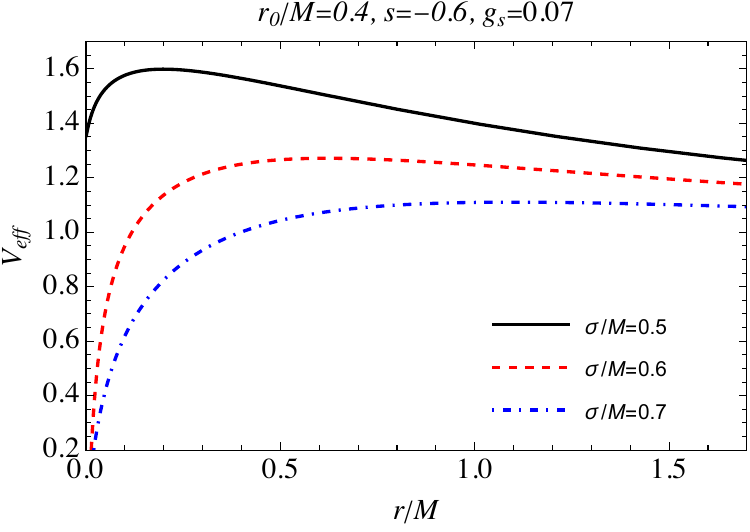}
\includegraphics[width=0.40\linewidth]{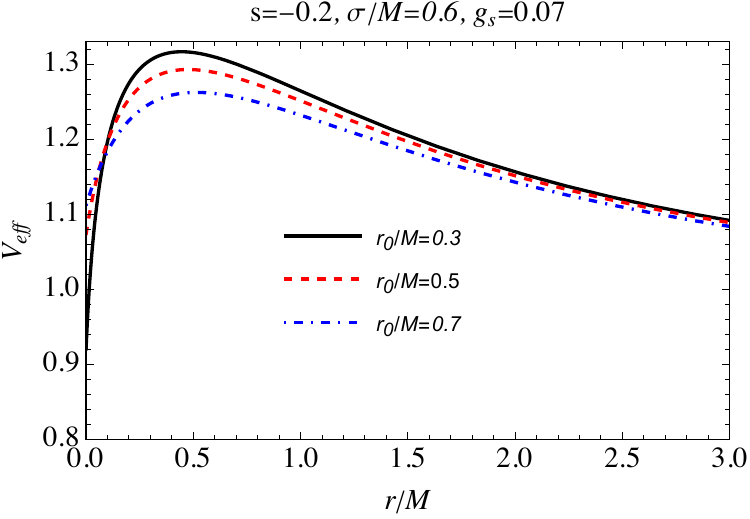}
\includegraphics[width=0.40\linewidth]{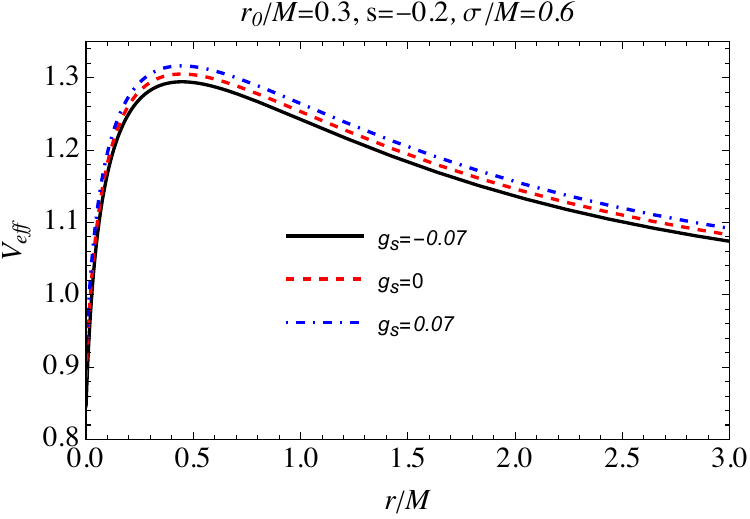}

\caption{ Effective potential $V_{\rm eff}$ of spinning test particles for different values of the spin parameter $s$, throat radius $r_0/M$, scalar field parameter $\sigma/M$, and coupling parameter $g_s$. \label{veff}}
\end{figure*}

Since only positive-energy trajectories are considered throughout this work, we use the positive branch of the effective potential,

\begin{align}
\mathcal{V}_{\rm eff}
=
\mathcal{V}_{\rm eff}^{+}.
\end{align}
Fig.~\ref{veff} shows the behavior of the effective potential for different values of the particle spin $s$, throat radius $r_0/M$, scalar field parameter $\sigma/M$, and scalar coupling parameter $g_s$. The height of the potential barrier increases with increasing values of the particle spin $s$ and the scalar coupling parameter $g_s$, whereas it decreases as the throat radius $r_0/M$ or the scalar field parameter $\sigma/M$ increases. Consequently, the model parameters substantially modify the radial motion of spinning particles and, through the extrema of the effective potential, influence the location and stability of the innermost stable circular orbit.

\subsection{Innermost stable circular orbit}
The circular orbits of spinning test particles are characterized by the following conditions. First, the radial velocity must vanish,

\begin{align}
\frac{dr}{d\lambda}=0,
\qquad
\mathcal{V}_{\rm eff}^{+}=\mathcal{E}.
\end{align}

Second, the radial acceleration must vanish,
\begin{align}
\frac{d^2r}{d\lambda^2}=0,
\qquad
\frac{d\mathcal{V}_{\rm eff}^{+}}{dr}=0.
\end{align}
These two conditions alone are not sufficient to guarantee the stability of circular motion. Stability requires

\begin{align}
\frac{d^2\mathcal{V}_{\rm eff}^{+}}{dr^2}>0.
\end{align}

The marginally stable circular orbit is obtained from the equality

\begin{align}
\frac{d^2\mathcal{V}_{\rm eff}^{+}}{dr^2}=0.
\end{align}

This orbit defines the innermost stable circular orbit (ISCO).

Fig. ~\ref{isco} presents the dependence of the ISCO radius $r_{\rm ISCO}$, angular momentum $\mathcal{L}_{\rm ISCO}$, and energy $\mathcal{E}_{\rm ISCO}$ on the particle spin for different values of the scalar coupling parameter $g_s$. The corresponding numerical values are listed in Table~\ref{tab:isco_gs}. For a fixed spin, increasing $g_s$ leads to larger values of $r_{\rm ISCO}$ and $\mathcal{E}_{\rm ISCO}$, whereas the corresponding angular momentum $\mathcal{L}_{\rm ISCO}$ decreases. In contrast, increasing the particle spin shifts the ISCO toward smaller radii. These results show that the scalar coupling has a significant influence on the properties of stable circular orbits around the wormhole.
\begin{widetext}
\begin{table*}[ht]
\centering
\caption{ISCO parameters for different values of the scalar coupling parameter $g_s$ and particle spin $s$.}
\label{tab:isco_gs}

\renewcommand{\arraystretch}{1.25}
\setlength{\tabcolsep}{10pt}

\resizebox{\textwidth}{!}{
\begin{tabular}{|c|c|c|c|c|c|c|c|c|c|}
\hline

$g_s$
& \multicolumn{3}{c|}{$s=-0.2$}
& \multicolumn{3}{c|}{$s=0$}
& \multicolumn{3}{c|}{$s=0.2$} \\
\cline{2-10}

&
$r_{\rm ISCO}$ & $\mathcal{L}_{\rm ISCO}$ & $\mathcal{E}_{\rm ISCO}$
&
$r_{\rm ISCO}$ & $\mathcal{L}_{\rm ISCO}$ & $\mathcal{E}_{\rm ISCO}$
&
$r_{\rm ISCO}$ & $\mathcal{L}_{\rm ISCO}$ & $\mathcal{E}_{\rm ISCO}$ \\
\hline

-0.05 & 2.65468 & 2.52489 & 0.942817 & 2.36523 & 2.42572 & 0.938534 & 2.03040 & 2.30953 & 0.932771 \\
-0.04 & 2.66285 & 2.51964 & 0.943339 & 2.37286 & 2.42084 & 0.939096 & 2.03759 & 2.30512 & 0.933390 \\
-0.03 & 2.67107 & 2.51435 & 0.943860 & 2.38054 & 2.41591 & 0.939657 & 2.04482 & 2.30067 & 0.934007 \\
-0.02 & 2.67934 & 2.50902 & 0.944379 & 2.38826 & 2.41095 & 0.940215 & 2.05208 & 2.29617 & 0.934622 \\
-0.01 & 2.68766 & 2.50365 & 0.944896 & 2.39603 & 2.40594 & 0.940772 & 2.05938 & 2.29164 & 0.935235 \\
 0.00 & 2.69603 & 2.49824 & 0.945412 & 2.40384 & 2.40089 & 0.941327 & 2.06672 & 2.28707 & 0.935845 \\
 0.01 & 2.70446 & 2.49279 & 0.945927 & 2.41170 & 2.39581 & 0.941881 & 2.07410 & 2.28245 & 0.936454 \\
 0.02 & 2.71295 & 2.48730 & 0.946439 & 2.41961 & 2.39068 & 0.942432 & 2.08152 & 2.27780 & 0.937060 \\
 0.03 & 2.72149 & 2.48177 & 0.946951 & 2.42757 & 2.38551 & 0.942982 & 2.08898 & 2.27310 & 0.937665 \\
 0.04 & 2.73009 & 2.47619 & 0.947461 & 2.43558 & 2.38030 & 0.943530 & 2.09648 & 2.26836 & 0.938268 \\
 0.05 & 2.73875 & 2.47058 & 0.947969 & 2.44364 & 2.37504 & 0.944077 & 2.10403 & 2.26358 & 0.938868 \\

\hline

\end{tabular}
}
\end{table*}
\end{widetext}

\subsection{Superluminal bound}
Before analyzing the motion of spinning particles, one must also take into account the superluminal bound. In the MPD formalism, the four-momentum $p^\alpha$ and the four-velocity $u^\alpha$ are generally not parallel. To quantify their relation, one introduces the effective mass
\begin{align}
    \mu=-p^\beta u_\beta.
\end{align}
Using eq~\eqref{MPD}, the four-momentum can be expressed as \cite{Ladino2023}
\begin{align}
    p^\alpha=\mu u^\alpha-u^\beta\frac{D S^{\alpha\beta}}{D\lambda},
\end{align}
which explicitly shows that the momentum and velocity vectors are not parallel for spinning particles. Consequently, although the momentum satisfies
\[
p^\alpha p_\alpha=-m^2(1+g_s\varphi)^2,
\]
the standard normalization of the four-velocity,
\[
u^\alpha u_\alpha=-1,
\]
is not automatically guaranteed.

For some values of the spin and radial coordinate, the trajectory may become spacelike. Such motion is unphysical for massive particles. Therefore, the timelike condition must be imposed. The limiting case is determined by
\begin{align}
u^\alpha u_\alpha=0.
\end{align}
Equivalently, the physical timelike region is specified by
\begin{align}
\frac{u^\alpha u_\alpha}{(u^t)^2}
=
g_{tt}
+
g_{rr}
\left(
\frac{dr}{dt}
\right)^2
+
g_{\phi\phi}
\left(
\frac{d\phi}{dt}
\right)^2
\leq 0 .
\label{eq:superluminal}
\end{align}

\begin{figure}
    \centering
    \includegraphics[width=0.8\linewidth]{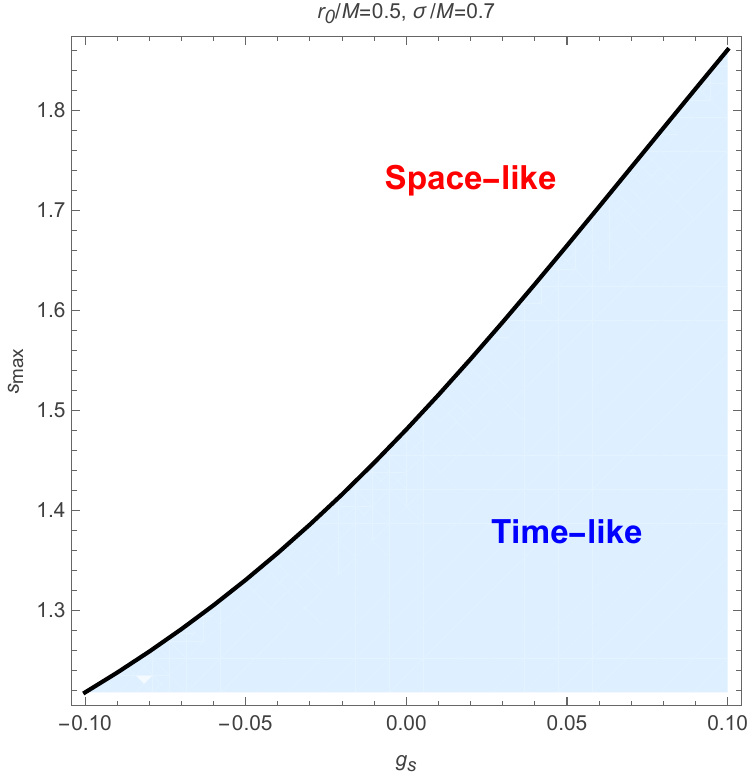}
    \caption{
Maximum allowed spin parameter $s_{\max}$ as a function of the coupling parameter $g_s$.}

    \label{smax}
\end{figure}

Solving the MPD equations gives

\begin{align}
\frac{dr}{dt}
=
\frac{u^r}{u^t}
=
\frac{Cp^r}{Bp^t},
\end{align}
and

\begin{align}
\frac{d\phi}{dt}
=
\frac{u^\phi}{u^t}
=
\frac{Ap^\phi}{Bp^t},
\end{align}
where

\begin{align}
A
&=
g_{\phi\phi}
+
\left(
\frac{S^{\phi r}}{p_t}
\right)^2
R_{trrt},
\\
B
&=
g_{tt}
+
\left(
\frac{S^{\phi r}}{p_t}
\right)^2
R_{t\phi\phi t},
\\
C
&=
g_{rr}
+
\left(
\frac{S^{\phi r}}{p_t}
\right)^2
R_{\phi t\phi t}.
\end{align}
As shown by the vertical lines in Fig.~\ref{isco}, the superluminal bound depends on the scalar coupling parameter $g_s$. These lines mark the critical spin values separating the physically allowed timelike region from the spacelike region. The corresponding dependence of the maximum allowed spin on the scalar coupling is displayed in Fig.~\ref{smax}, and the numerical values are listed in Table~\ref{tab:smax}. It is clear that increasing $g_s$ shifts the critical spin to higher values, thereby extending the range of physically admissible spinning particle motion before the onset of superluminal behavior.
\begin{table}[ht]
\centering
\caption{Maximum allowed spin parameter $s_{\max}$ for different values of $g_s$.}
\renewcommand{\arraystretch}{1.4}

\begin{tabular}{c@{\hspace{1cm}}c}

\begin{tabular}{cc}
\hline
$g_s$ & $s_{\max}$\\
\hline
-0.09 & 1.23746\\
-0.08 & 1.25854\\
-0.07 & 1.28097\\
-0.06 & 1.30482\\
-0.05 & 1.33017\\
-0.04 & 1.35708\\
-0.03 & 1.38559\\
-0.02 & 1.41570\\
-0.01 & 1.44739\\
 0.00 & 1.48059\\
\hline
\end{tabular}

&

\begin{tabular}{cc}
\hline
$g_s$ & $s_{\max}$\\
\hline
0.01 & 1.51521\\
0.02 & 1.55109\\
0.03 & 1.58807\\
0.04 & 1.62595\\
0.05 & 1.66450\\
0.06 & 1.70351\\
0.07 & 1.74278\\
0.08 & 1.78212\\
0.09 & 1.82135\\
0.10 & 1.86033\\
\hline
\end{tabular}

\end{tabular}

\label{tab:smax}
\end{table}

\begin{figure*}\centering
\includegraphics[width=0.30\linewidth]{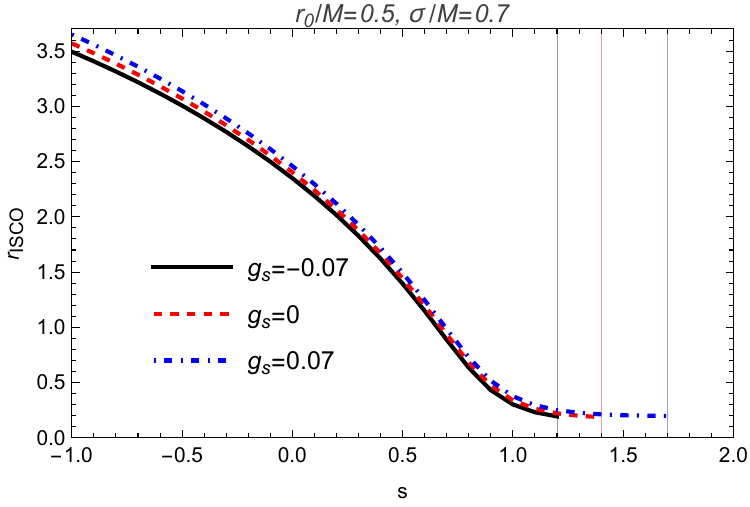}
\includegraphics[width=0.30\linewidth]{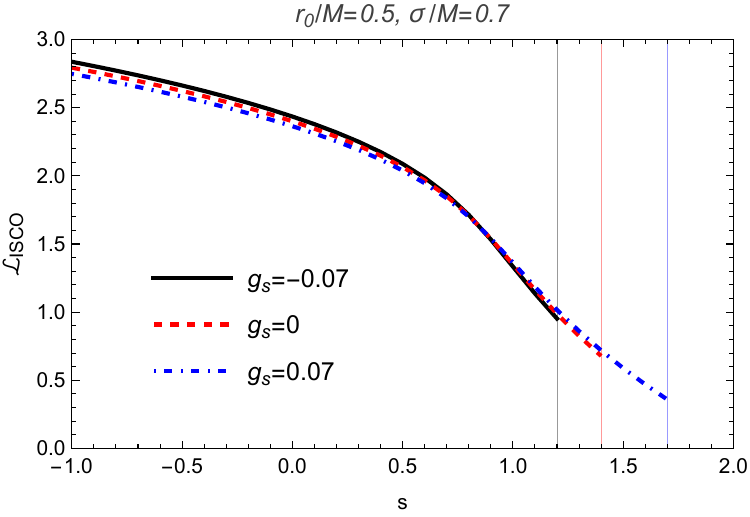}
\includegraphics[width=0.30\linewidth]{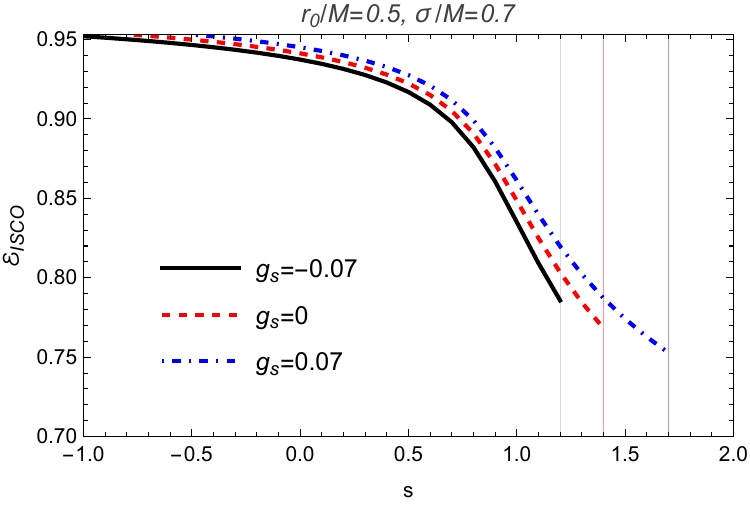}

\caption{
ISCO radius $r_{\rm ISCO}$, angular momentum $L_{\rm ISCO}$, and energy $E_{\rm ISCO}$ as functions of the particle spin parameter $s$ for $g_s=-0.07$, $0$, and $0.07$. The throat radius and scalar field parameter are fixed at $r_0/M=0.5$ and $\sigma/M=0.7$, respectively. The vertical lines denote the critical spin $s_{\max}$ separating the physically admissible timelike region from the spacelike region.
}
\label{isco}
\end{figure*}
\begin{figure*}
\centering
\includegraphics[width=0.40\linewidth]{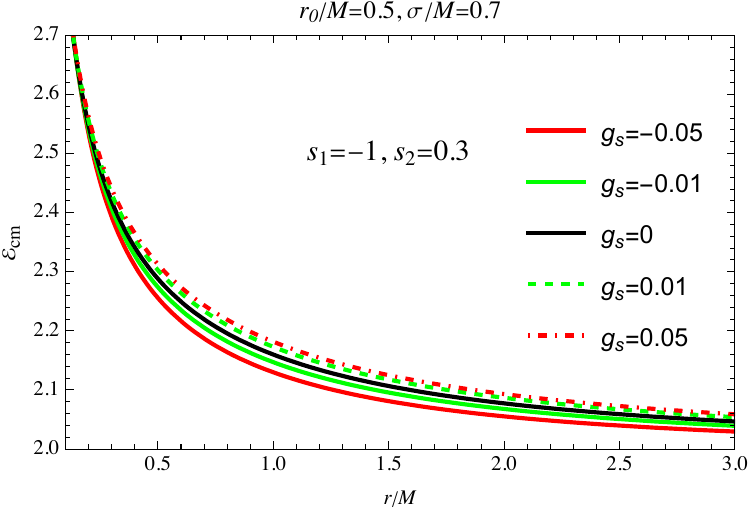}
\includegraphics[width=0.40\linewidth]{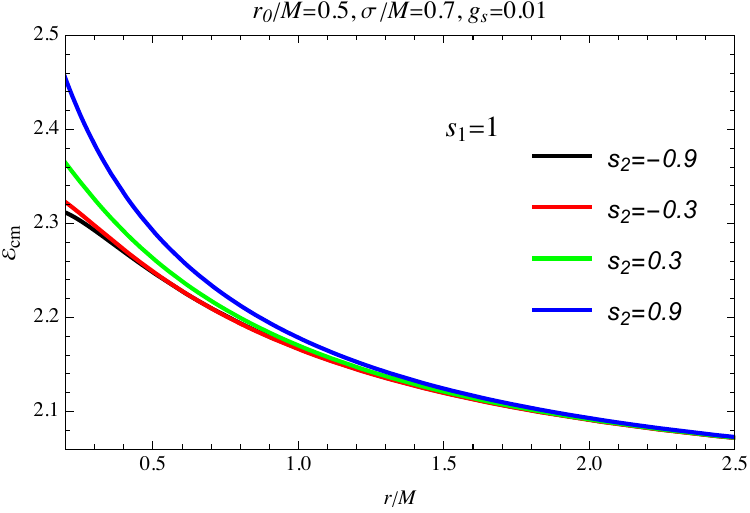}
\includegraphics[width=0.40\linewidth]{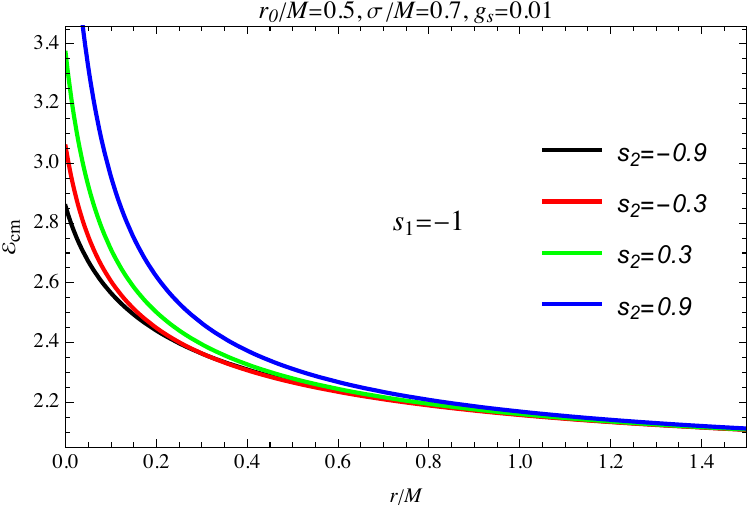}
\includegraphics[width=0.40\linewidth]{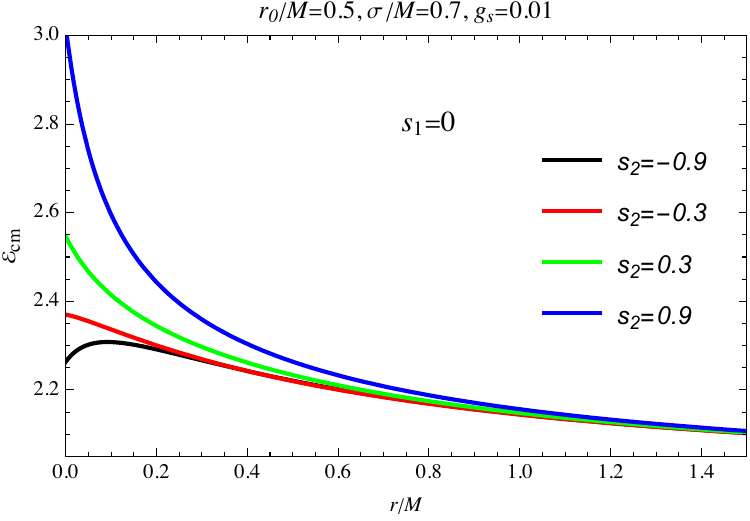}
\caption{
Dimensionless center-of-mass energy $\mathcal{E}_{\rm cm}$ as a function of the radial coordinate $r/M$. The upper-left panel shows the dependence on the scalar-field coupling parameter $g_s$, while the remaining panels illustrate the effect of the spin parameter $s_2$ for fixed values of $s_1$.
}
\label{ecm}
\end{figure*}

\subsection{Center-of-mass energy}

To investigate the influence of spin--curvature coupling on particle collisions in the vicinity of the compact object, we evaluate the center-of-mass (CM) energy of two colliding spinning particles. The center-of-mass energy represents the total energy available in the local rest frame of the collision and serves as an important quantity for studying high-energy collision processes in strong gravitational fields.

Following the standard definition, the center-of-mass energy is given by

\begin{eqnarray}
E_{cm}^2
&=&
-g^{\mu\nu}
\left(p^{(1)}_{\mu}+p^{(2)}_{\mu}\right)
\left(p^{(1)}_{\nu}+p^{(2)}_{\nu}\right)
\nonumber\\
&=&
m_1^2+m_2^2
-
2g^{\mu\nu}
p^{(1)}_{\mu}
p^{(2)}_{\nu},
\label{Ecm}
\end{eqnarray}

where $p^{(1)}_{\mu}$ and $p^{(2)}_{\mu}$ denote the covariant four-momenta of the first and second spinning particles, respectively.

For simplicity, we consider two particles with identical rest masses, $m_1=m_2=m$, while allowing their four-momenta and conserved quantities to be different. Throughout this analysis, the particles are assumed to move in opposite azimuthal directions with opposite orbital angular momenta, so that they undergo a head-on collision in the equatorial plane. Under these assumptions, the dimensionless center-of-mass energy can be written as

\begin{eqnarray}
\mathcal{E}_{cm}^2
&=&
\frac{E_{cm}^2}{2m^2 (1+g_s \varphi)^2}
\nonumber\\
&=&
1-
\frac{
g^{tt}p_t^{(1)}p_t^{(2)}
+
g^{rr}p_r^{(1)}p_r^{(2)}
+
g^{\phi\phi}
p_{\phi}^{(1)}
p_{\phi}^{(2)}
}
{m^2 (1+g_s \varphi)^2}.
\label{eq:Ecm_dimensionless}
\end{eqnarray}

The radial momentum components entering the above expression are determined from the corresponding equations of motion, whereas the temporal and azimuthal momentum components are obtained from the conserved energy and angular momentum of each spinning particle. Consequently, the center-of-mass energy depends not only on the background spacetime geometry but also on the particle spin through the spin--curvature interaction. In the spinless limit ($s_1=s_2=0$), the above expression naturally reduces to the familiar center-of-mass energy for two geodesic test particles.

Fig. ~\ref{ecm} illustrates the radial dependence of the dimensionless center-of-mass energy $\mathcal{E}_{\rm cm}$ for different values of the scalar coupling parameter $g_s$ and the particle spins. In all panels, the collision energy reaches its maximum close to the wormhole throat and decreases monotonically with increasing radial distance.

The upper-left panel shows that increasing the scalar coupling parameter $g_s$ systematically enhances the center-of-mass energy over the entire radial range. The remaining panels demonstrate that the collision energy is also highly sensitive to the spin configuration of the colliding particles. For fixed $s_1$, the center-of-mass energy increases with increasing $s_2$. Moreover, configurations with negative values of $s_1$ produce noticeably larger collision energies than those with positive $s_1$, with the highest values obtained for $s_1=-1$ and $s_2=0.9$. These results indicate that both the scalar coupling and the particle spin configuration play important roles in determining the efficiency of high-energy particle collisions around the wormhole.

\section{Conclusion \label{sec6}}

In this work, we have investigated the dynamics of massive and spinning test particles interacting with a background scalar field in a three-parameter scalarized wormhole spacetime. The coupling between the scalar field and the test particles modifies their trajectories, the properties of stable circular orbits, oscillatory motion, and the energetics of particle collisions.

For spinless massive particles, we derived the effective potential together with the conserved energy and angular momentum, and analyzed the properties of the innermost stable circular orbit (ISCO). Our results show that increasing the scalar coupling shifts the ISCO to larger radii, implying that stronger scalar interactions move stable circular orbits farther from the wormhole throat. The orbital and radial epicyclic frequencies were then obtained by considering small perturbations about circular motion. We found that the scalar coupling generally suppresses both characteristic frequencies, whereas the wormhole throat radius and scalar charge influence the oscillatory dynamics in different ways. These modifications are directly reflected in the ER3 and ER4 epicyclic resonance models, where the location of the characteristic 3:2 resonance systematically shifts toward larger radii as the scalar coupling and wormhole parameters increase. This behavior suggests that QPO observations could, in principle, provide a useful observational probe of scalarized wormhole geometries.

The dynamics of spinning particles were investigated within the framework of the Mathisson–Papapetrou–Dixon equations supplemented by the Tulczyjew spin supplementary condition. We derived the corresponding effective potential and examined the influence of spin–curvature coupling on circular motion. Our analysis demonstrates that the particle spin significantly modifies both the effective potential and the ISCO properties. In particular, increasing the scalar coupling leads to larger values of the ISCO radius and the corresponding orbital energy, while the required angular momentum decreases. We also determined the superluminal bound associated with the non-parallelism of the particle four-momentum and four-velocity. The maximum physically admissible spin increases monotonically with the scalar coupling parameter, indicating that stronger scalar interactions enlarge the region of parameter space in which physically acceptable timelike trajectories exist.

Finally, we investigated the center-of-mass energy of particle collisions in the vicinity of the wormhole throat for both spinless and spinning particles. The collision energy reaches its maximum near the throat and decreases monotonically with increasing radial distance. For spinning particles, the collision energy depends not only on the scalar coupling but also on the relative spin orientation of the colliding particles. Our results show that anti-aligned spin configurations generate considerably higher center-of-mass energies than aligned configurations, demonstrating that the relative spin orientation plays a crucial role in determining the efficiency of particle acceleration in scalarized wormhole spacetimes.

Overall, our results demonstrate that the combined effects of the scalar field and spin–curvature interaction have a significant impact on orbital dynamics, epicyclic oscillations, and high-energy particle collisions around scalarized wormholes. These effects may provide potentially observable signatures capable of distinguishing scalarized wormholes from other compact objects. Future work will extend the present analysis to rotating scalarized wormholes, more general forms of scalar-field coupling, and detailed comparisons with observational QPO measurements and other astrophysical probes.

\bibliographystyle{apsrev4-1}
\bibliography{reference}

\end{document}